\documentclass[twocolumn]{emulateapj}
\usepackage{graphicx}
\usepackage{float}
\usepackage{mathrsfs}	
\usepackage{amsmath}
\usepackage{color}
\usepackage{multirow}
\usepackage{subfigure}
\usepackage{longtable}
\usepackage{blindtext}
\usepackage{amssymb}
\usepackage{lineno}

\def\gsim{~\rlap{$>$}{\lower 1.0ex\hbox{$\sim$}}}
\def\lsim{\mathrel{\rlap{\lower3.5pt\hbox{\hskip0.5pt$\sim$}}
    \raise0.5pt\hbox{$<$}}} 
\newcommand{\rui}[1]{ {\color{black}}}
\begin{document}
\newcommand{\chiara}[1]{{\color{magenta}{#1}}}

\title{High-quality strong lens candidates in the final Kilo-Degree survey footprint}

\author{\mbox{R. Li\altaffilmark{1,2,3}}}
\author{\mbox{N. R. Napolitano\altaffilmark{1,4,5}}}
\author{\mbox{C. Spiniello\altaffilmark{6,5}}}
\author{\mbox{C. Tortora\altaffilmark{5}}}
\author{\mbox{K. Kuijken\altaffilmark{7}}}
\author{\mbox{L. V. E. Koopmans\altaffilmark{8}}}
\author{\mbox{P. Schneider\altaffilmark{9}}}
\author{\mbox{F. Getman\altaffilmark{5}}}
\author{\mbox{L. Xie\altaffilmark{1}}}
\author{\mbox{L. Long \altaffilmark{1}}}
\author{\mbox{W. Shu \altaffilmark{1}}}
\author{\mbox{G. Vernardos\altaffilmark{8,10}}}
\author{\mbox{Z. Huang\altaffilmark{1}}}
\author{\mbox{G. Covone\altaffilmark{5,11,12}}}
\author{\mbox{A. Dvornik \altaffilmark{13}}}
\author{\mbox{C. Heymans\altaffilmark{14,13}}}
\author{\mbox{H. Hildebrandt\altaffilmark{13}}}
\author{\mbox{M. Radovich\altaffilmark{15}}}
\author{\mbox{A.H. Wright\altaffilmark{13}}}

\affil{\altaffilmark{1}School of Physics and Astronomy, Sun Yat-sen University, Zhuhai Campus, 2 Daxue Road, Xiangzhou District, Zhuhai, P. R. China {\it \rm napolitano@mail.sysu.edu.cn}}

\affil{\altaffilmark{2}School of Astronomy and Space Science, University of Chinese Academy of Sciences, Beijing 100049, China}

\affil{\altaffilmark{3}National Astronomical Observatories, Chinese Academy of Sciences, 20A Datun Road, Chaoyang District, Beijing 100012, China;}

\affil{\altaffilmark{4} CSST Science Center for Guangdong-Hong Kong-Macau Great Bay Area, Zhuhai, China, 519082}

\affil{\altaffilmark{5}INAF -- Osservatorio Astronomico di Capodimonte, Salita Moiariello 16, 80131 - Napoli, Italy }
\affil{\altaffilmark{6}Department of Physics, University of Oxford, Denys Wilkinson Building, Keble Road, Oxford OX1 3RH, UK}
\affil{\altaffilmark{7}Leiden Observatory, Leiden University, P.O.Box 9513, 2300RA Leiden, The Netherlands}
\affil{\altaffilmark{8}Kapteyn Astronomical Institute, University of Groningen, P.O.Box 800, 9700AV Groningen, {the Netherlands}}
\affil{\altaffilmark{9}Argelander-Institut für Astronomie, Auf dem Hügel 71, 53121 Bonn / Germany}
\affil{\altaffilmark{10}institute of Physics, Laboratory of Astrophysics, Ecole Polytechnique Fédérale de Lausanne (EPFL), Observatoire de Sauverny, 1290 Versoix, Switzerland}
\affil{\altaffilmark{11}Dipartimento di Fisca ``E. Pancini'', University of Naples ``Federico II'', Naples, Italy}
\affil{\altaffilmark{12}INFN, Sezione di Napoli,  Naples, Italy}
\affil{\altaffilmark{13}Ruhr University Bochum, Faculty of Physics and Astronomy, Astronomical Institute (AIRUB), German Centre for Cosmological Lensing, 44780 Bochum, Germany}
\affil{\altaffilmark{14}Institute for Astronomy, University of Edinburgh, Royal Observatory, Blackford Hill, Edinburgh, EH9 3HJ, UK}
\affil{\altaffilmark{15}INAF - Osservatorio Astronomico di Padova, via dell'Osservatorio 5, 35122 Padova, Italy}

\begin{abstract}
We present 97 new high-quality strong lensing candidates found 
in the final $\sim 350\,\rm deg^2$, that completed the full $\sim 1350\,\rm deg^2$ area of the Kilo-Degree Survey (KiDS).
Together with our previous findings, the final list of  high-quality candidates from KiDS sums up to 268 systems.
The new sample is assembled using a new Convolutional Neural Network (CNN) classifier applied to $r$-band (best seeing) and $g,~r,~i$ color-composited images separately. This optimizes the complementarity of the morphology and color information on the identification of strong lensing candidates.  We apply the new classifiers to a sample of luminous red galaxies (LRGs) and a sample of bright galaxies (BGs) and select candidates that received a high probability to be a lens from the CNN ($P_{\rm CNN}$). In particular, 
setting $P_{\rm CNN}>0.8$ for the LRGs, the $1$-band CNN predicts 1213 candidates, while the $3$-band classifier yields 1299 candidates, with only $\sim$30\% overlap. 
For the BGs, in order to minimize the false positives, we adopt a more conservative threshold, $P_{\rm CNN} >0.9$, for both CNN classifiers. This results in 3740 newly selected objects. The candidates from the two samples are visually inspected by 7 co-authors to finally select 97 ``high-quality'' lens candidates which received mean scores larger than 6 (on a scale from 0 to 10). We finally discuss the effect of the seeing on the accuracy of CNN classification and possible avenues to increase the efficiency of multi-band classifiers, in preparation of next-generation surveys from ground and space.
\end{abstract}

\section{Introduction}
\label{sec:intro}
Strong gravitational lensing (SGL, hereafter) is the effect of deformation of the images of 
distant sources,
induced by the gravitational field of massive, typically red and dead, galaxy lenses
located along the line-of-sight of the observer to the source. 
According to general relativity, the light from the source 
is magnified and deflected by the curved space-time generated by the lens, or deflector, to form distinctive SGL features. 
If the source is a compact object 
like a quasar (e.g., \citealt{2013ApJ...766...70S, 2017MNRAS.468.2590S}) a supernova (e.g., \citealt{2015Sci...347.1123K})  or an ultra-compact galaxy (e.g., \citealt{2006ApJ...646L..45B, 2017MNRAS.465.2411M, 2020ApJ...904L..31N}) its light is split into multiple images. If  
the source is an extended galaxy, it produces a stretched arc or full ring (e.g., \citealt{2008ApJ...682..964B, 2012ApJ...744...41B, 2013ApJ...777...98S}). 

SGL is a powerful tool to infer both the distribution of dark matter (DM) of the lenses and the properties of high-redshift sources. Specifically, SGL allows one to measure the mass of the deflectors with much higher accuracy than any other methods (e.g., \citealt{2006ApJ...649..599K, 2009ApJ...703L..51K, 2009ApJ...705.1099A, 2010ApJ...724..511A, 2012ApJ...757...82B, 2015ApJ...803...71S,2018MNRAS.480..431L}). SGL can further be used to constrain DM substructures (e.g., \citealt{2012Natur.481..341V, 2017MNRAS.468.1426L,2020MNRAS.492.3047H}) and the expansion history of the Universe (e.g., \citealt{2013ApJ...766...70S, 2017MNRAS.468.2590S, 2017MNRAS.465.4914B, 2019MNRAS.490..613S}). 
Besides cosmology-related questions, SGL can also be used for galaxy formation and evolution studies. 
Indeed, acting as a gravitational telescope, it magnifies high-redshift objects, which otherwise would be difficult to observe, hence permitting the study of their internal structure and stellar populations (e.g., \citealt{2015ApJ...808L...4A, 2018ApJ...853..148C, 2019MNRAS.489.5022C, 2019ApJ...881....8C, 2020MNRAS.491.2447R, 2020ApJ...904L..31N}).

Unfortunately, SGL events are {rare, }
since the strong lensing cross-section subtends only a small angular region around the lens center, typically a few tenths of an arcsec. 
For instance, we expect to find $\sim 0.5$ arcs (\citealt{2015ApJ...811...20C}) or $\sim 0.025$ quadruple images per deg$^2$ (\citealt{2010MNRAS.405.2579O}) in typical ground-based surveys with $r$-band limiting AB magnitudes $\sim 25$. 
Hence, only by mapping large portions of the sky {with deep multi-band photometric surveys,} it is possible to build statistically large
lens samples and span a large variety of deflectors and sources.

This opportunity is offered by
ongoing programs like the Kilo-Degree Survey (KiDS, \citealt{2013Msngr.154...44D}), the Hyper Suprime-Cam survey (HSC, \citealt{2012SPIE.8446E..0ZM}), and the Dark Energy Survey (DES, \citealt{2005astro.ph.10346T}), which are providing large databases of galaxies including millions of objects and thousands of lens candidates (e.g., \citealt{2021ApJ...909...27H}). 

Next generation sky surveys (e.g., Vera Rubin/LSST survey, \citealt{2019ApJ...873..111I}; the {\it Euclid} mission, \citealt{2011arXiv1110.3193L}; the Chinese Space Station Telescope, CSST, \citealt{2018cosp...42E3821Z}) will offer 
even larger databases of billions of objects in the  
near future, where we expect to find hundreds of thousands of SGLs (see e.g., \citealt{2015ApJ...811...20C}).
The drawback is that the search for rare strong lensing events in such a gigantic number of galaxies cannot be done by visual inspection, but needs specialized 
tools.  

Luckily,
the combination of characteristic morphological patterns and different colors of lens and source images are distinctive features {that can be used to identify these rare treasures among millions of galaxies.}
This feature recognition is a classical application for machine learning techniques, which can typically handle it with small computational times. 
For this reason, in the past years, a large effort has been made to develop machine learning-based algorithms to search for lenses in large sky surveys (e.g., \citealt{2015MNRAS.448.1446A, 2017MNRAS.465.4325O, 2019A&A...632A..56K, Speagle2019,2020ApJ...894...78H}). In particular,  Convolutional Neural Networks (CNNs) have proven to be very effective in finding lenses from imaging (e.g. \citealt{2019ApJS..243...17J, 2019MNRAS.484.3879P, 2020ApJ...899...30L}) and spectroscopic data (\citealt{2019MNRAS.482..313L}).

Indeed, many high-quality/reliability lens candidates have been already found among the higher-ranked candidates from CNNs by human inspection (see e.g., \citealt{2020ApJ...899...30L}, also \citealt{2019A&A...625A.119M} for a discussion), and a subsample of them has already been spectroscopically confirmed (see e.g., \citealt{2019MNRAS.483.3888S}, \citealt{2019MNRAS.485.5086S} \citealt{2020MNRAS.494.3491L}, \citealt{2020MNRAS.494.1308N}, \citealt{2020ApJ...904L..31N}).
Hence, whereas CNNs are formidable tools to provide reliable candidates in the newly explored areas of the sky for follow-up observations,
further improvement of CNN methods are needed to maximize the completeness and purity of the candidate lists (see e.g., \citealt{2019MNRAS.484.3879P}), especially for medium/low signal-to-noise ratio arcs and multiple images. 

In this work, we continue our effort, started in \citet{2017MNRAS.472.1129P}, to develop CNN algorithms to identify strong gravitational lens candidates in high-quality multi-band optical images from the KiDS survey \citep{2013Msngr.154...44D}. We make significant improvements both in the CNN architecture and in the preparation of its training sample. In \citet{2019MNRAS.484.3879P},  
\citep[][P+19 hereafter]{2019MNRAS.482..807P} and \citep[][L+20 hereafter]{2020ApJ...899...30L}, we demonstrated that CNN classifiers can identify known lenses, and produced lists of high-quality SGL candidates from all publicly available KiDS Data Releases (DRs).

In this paper, we extend the search to the internal Data Release 5 (KiDS-iDR5).
This is a partial data release, only made available to the KiDS consortium for science-driven tests. It includes the stacked images of 341 tiles in $u,g,r,i$ optical bands, the weight maps, and the optical multi-band catalogs, containing aperture photometry of all detected sources in each tile. These are meta-products that are routinely produced by KiDS since the first data release (\citealt{2015A&A...582A..62D}). The data reduction and calibration, as well as the image astrometry and photometry, are obtained with the same procedures of 
Data Release 4 (KiDS-DR4) described in \citet{2019A&A...625A...2K}. Observational constraints and image quality are comparable to KiDS-DR4, while no photometric redshifts are available as the 9-band catalogs including Gaussianized aperture photometry (see \citealt{2019A&A...625A...2K}) are still under production. The official DR5 release will also contain multi-epoch imaging in the $i$-band, which is not yet available for the iDR5 and will be investigated in future analyzes.

Despite a number of new features implemented in the new classifier, we found some modules that need to be optimized, in particular, to improve the completeness and purity of the candidates.
For this reason, our primary objective in this paper is to present a catalog of {very high-quality (HQ)} new strong lens candidates found in a yet unexplored sky area.
{We aim at providing a reference catalog to be used for the target selection of forthcoming spectroscopic surveys (e.g. 4MOST, \citealt{2019Msngr.175....3D}),} although we are aware that this new sample might not be fully complete {nor entirely pure}.

The paper is organized as follows. In Sect.\,2, we 
describe {the datasets used to }
train and test the classifiers. In Sect.\,3, we 
apply our CNN classifiers to two predictive datasets, the luminous red galaxies (LRGs), and the bright galaxies (BGs) and discuss their performance.
We present the new lens candidates and discuss the possible avenues to further improve
the accuracy of the classifiers in Sect.\,4, and summarize our main conclusions in Sect.\,5.

\section{New Convolutional Neural Network classifier}
\label{sec:new_classifier}
\subsection{The CNN classifier}
\label{sec:CNN_classifier}

CNNs use convolution kernels as artificial neurons to capture the local features of input images. This makes them particularly suitable for feature recognition and hence for the identification of strong gravitational lenses.

In this work, we use ResNet (\citealt{2015arXiv151203385H}). This is one of the most widely used CNN models in lens search analyzes, where it has been found to perform very efficiently  (e.g., \citealt{2017AAS...22934205L, 2019MNRAS.484.3879P, 2019MNRAS.482..807P}). Different from our previous work (\citealt{2020ApJ...899...30L}), where we used the open-source keras-resnet\footnote{https://github.com/raghakot/keras-resnet} code, we here customize a new architecture with Keras\footnote{https://github.com/keras-team/keras} running on the backend of TensorFlow\footnote{https://github.com/tensorflow/tensorflow}. 

We build up two different classifiers, although with the same architecture and the same number of  convolutional layers (18). The first one (1-band CNN hereafter) uses only $r$-band images and thus searches for lens candidates based on morphological patterns. The second (3-band CNN) uses instead color-composite $g$, $r$, $i$ images and hence is also sensitive to color contrast between the deflector and the source. 
In both cases, we use images of $101 \times 101$ pixels (corresponding to $20''\times20''$) as input, and we obtain the probability of a system to be a lensing event, $P_{\rm CNN}$, as output. 

A final technical note is related to the 
{\it data augmentation}. This is a standard strategy used in CNNs to avoid over-fitting and to improve the robustness of the classifier. As we will discuss in the next section, during the training of the CNN we
provide a fresh supervised sample of ``features'', i.e. images of true and false SGL events, and ``labels'', i.e. true or false lens, {at every training step}. 
To virtually increase the training sample and generalize the features, we have customized the
data augmentation, including randomly shift, flip, rotation, crop, and color vibrance (saturation, contrast, brightness, sharpness) of the images inputted to the CNN during the training phase. 

\subsection{The training and testing data}
\label{sec:training_data}
When building a CNN classifier, the principal step is network {\em training}. In this phase the CNN learns how to distinguish positive detections (``positives'', in short), i.e. galaxies with {\em true} lensing features, from
negative detections (``negatives'', hereafter), i.e. contaminants.
These latter are galaxies with features that can mimic gravitational arcs, such as spiral arms, polar rings, interactive systems, etc.

\begin{figure*}
    \centering
    \includegraphics[width=17cm]{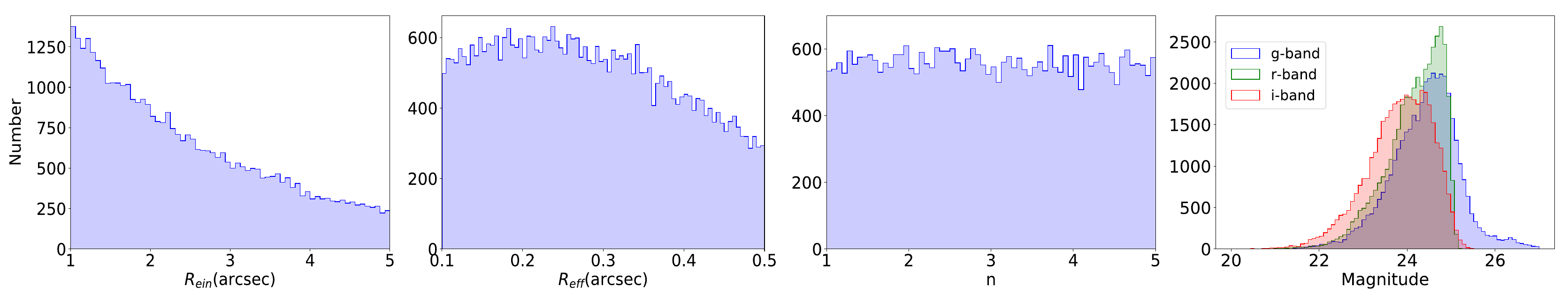}\vspace{3pt}
    \setlength{\abovecaptionskip}{0.0cm}
    \caption{Distribution of {the main }
    input parameters in the lensing simulation. From left to right {we plot }
    the Einstein radius, {the } source effective radius, {the } source S{\'e}rsic index and the $3$-band ($g,r,i$) source magnitudes. See text for details.}
    \label{fig:parameter_distribution}
\end{figure*}

A proper CNN training process requires the use of tens of thousands of positives and negatives with observational properties (e.g. seeing, noise, depth, etc) similar to the objects to be classified. However, even using all the existing confirmed lenses, their number would still be insufficient for this purpose.

To overcome this problem, a common approach is to produce 
mock positives. This can be done either by obtaining mock observations of fully simulated systems via ray-tracing of galaxies in hydrodynamical simulations (\citealt{2018MNRAS.473.3895L, 2020MNRAS.497..556H}) or by adding simulated arcs to {images of real} galaxies (e.g. \citealt{ 2019MNRAS.484.3879P,2020A&A...644A.163C}). Here we 
follow this latter approach using KiDS images of galaxies to which we add simulated arcs, as already done in L+20. In this case, the quality of the training images matches that of the input images and the deflectors reproduce the real foreground light, in terms of color, size ,and shape. 

For the positives, we simulated 43\,000 mock lens systems starting from 20\,000 KiDS DR4 images of LRGs and randomly adding to PSF-convolved simulated arcs/point images. 
The LRG sample was visually inspected to avoid galaxies that could have arc-like features around, mimicking a lensing event, as well as galaxies with poor image quality or affected by stellar diffraction spikes or reflection haloes.

The same LRGs are also used as negatives, together with other 23\,000 bright galaxies of every kind, selected from KiDS DR4 to be brighter than  $r=21$. Adding the positives and the negatives defined above, we obtain a total sample of 86\,000 objects that we need to split in the training and test samples. 80\,000 (40\,000 positives plus 40\,000 negatives) are used to train the CNNs, and 6000 (3000 positives plus 3000 negatives) are used to test the performance.



To simulate positives, we start by creating the multi-band arcs and multiple images in $g,~r,~i$ bands with a ray-tracing method. We assume a S{\'e}rsic profile for {the} sources and a singular isothermal ellipsoid (SIE) mass model for the deflectors. 
We use the color information of the Rubin/LSST mock galaxies catalog \citep{2010SPIE.7738E..1OC} to build a suitable source color library to
obtain realistic colors of lensed images. 
This catalog has a photometric depth of $r\sim 28$ and covers the redshift range $0<z<6$. It was generated from the Millennium Simulation \citep{2005Natur.435..629S}, with superimposed galaxies based on a semi-analytical model for galaxy evolution (\citealt{2006MNRAS.366..499D}) which includes gas cooling, star formation, and supernovae/AGN feedback, to reproduce the observed colors, luminosities, and clustering of galaxies. We select
$\sim 2600$ of these mock galaxies at redshift between 0.8 and 3 and with $r$-band {AB} magnitudes between 21 and 25. When building the ray-tracing images,
the $g,~r,~i$ band magnitudes of the S{\'e}rsic sources are randomly selected from this color library. The parameters of the SIE lens and S{\'e}rsic source profiles (e.g. Einstein radius, effective radius, S\'ersic index, axis ratio, etc.),
used to simulate the lensed images, are randomly generated from probability distributions, as reported in Table \ref{tab:parameters}. For the parameter distributions, we have followed
P+19 and L+20.
The distributions of the Einstein radius ($R_{\rm ein}$) and the source effective radius ($R_{\rm eff}$) are exponential and normal, respectively. 
The distribution of the S{\'e}rsic index ($n$),  
position angle and ellipticity of the lenses and sources are originally uniform. However, as we will describe below, we apply a further selection on signal-to-noise ratio and relative brightness of the lens galaxies and lensed images.
We also perturb the magnitudes in all three bands by randomly adding a value between $\pm 0.1$ to account for some scatter around the nominal Rubin/LSST mock colors. 
The final distribution for  $R_{\rm ein}$, $n$ and $R_{\rm eff}$, together with the magnitude distribution in the three optical bands used by the CNNs
are shown in Fig.~\ref{fig:parameter_distribution}.

The next step is to obtain lensed images in each band, and to convolve them 
with a simulated point spread function (PSF), assumed to have a Moffat profile. The range and distribution of the full width at half maximum (FWHM) of the adopted PSFs are chosen according to the seeing distribution of KiDS \citep{2019A&A...625A...2K} and are shown in Table \ref{tab:parameters}. 
This is a significant improvement over the use of an average PSF for all tiles,
adopted in previous analyzes. In fact, 
the PSF variation leads to a large variance in the sharpness of the lensing features (see
Sect.\,\ref{subsec:discussion_of_performance} for a detailed discussion). This might not particularly impact the $r$-band, which is the best image quality filter in KiDS with a 
narrow FWHM distribution ($<0.8''$), but it can strongly affect the classification in the $3$-band CNN, since the seeing in the $g$ and $i$-bands can be significantly worse than that of the $r$-band images. Hence, 
by accounting for the FWHM variance in all filters, we can improve the {performance and } robustness of the {$3$-band} classifier. 

Finally, as done in L+20,
we also add 
(1) {an} 
external shear to account for the effect of different environments, and 
{(2) simulated Gaussian random field (GRF) fluctuations with a power-law power spectrum to the lens potential to account for the effect of the sub-halos in the lens plane (\citealt{2018MNRAS.474.1762C}). The slope of the power-law is fixed to $-6$ while the amplitude is determined using Parseval’s theorem, which is related to the variance of the GRF potential fluctuations inside the image via a normalization factor. Here, the variance for determining the amplitude is drawn by a logarithmic distribution between $10^{-4}$ -- $10^{-1}$, about mean zero in the units of the square of the lensing potential (see also in P+19, and detail description in \citealt{2018MNRAS.474.1762C}). This yields both structured sources and lenses that are not perfect SIE.}

\begin{deluxetable}{llcl}
\tablewidth{0pt}
\tablehead{
Parameter & Range &units& Distribution
}
\startdata
\hline
\multicolumn{4}{c}{lens (SIE)}\\ 
\hline
Einstein radius & 1.0 -- 5.0 &arcsec & exponential\\
Axis ratio & 0.4 -- 1.0 &--& uniform\\
Position angle & 0 -- 180 &degree& uniform\\
External shear & 0 -- 0.1 &--& uniform\\
Angle of external shear & 0 -- 180 &degree& uniform\\
\hline
\multicolumn{4}{c}{Source (S{\'e}rsic)}\\ 
\hline
Effective radius & 0.1 -- 0.5 &arcsec& normal\\
 &  && ($\mu=0.2$, $\sigma=0.3$)\\
Axis ratio & 0.3 -- 1.0 &--& uniform\\
Position angle & 0 -- 180 &degree& uniform\\
S{\'e}rsic index & 0.3 -- 5.0 &--& uniform\\
\hline
\multicolumn{4}{c}{PSF (Moffat)}\\ 
\hline
FWHM-$g$ & 0.60 -- 1.2 &--& normal\\
 &  && ($\mu=0.85$, $\sigma=0.1$)\\
FWHM-$r$ & 0.50 -- 0.90 &--& normal\\
 &  && ($\mu=0.7$, $\sigma=0.05$)\\
FWHM-$i$ & 0.55 -- 1.2 &--& normal\\
 &  && ($\mu=0.80$, $\sigma=0.1$)\\
$\beta$ & 2.20 && fixed\\
Axis ratio & 0.98 -- 1.02 &--& uniform\\
Position angle & 0 -- 180 &--& uniform
\enddata
\tablecomments{Range and distribution of parameter values used to simulate the lensed images. {$\beta$ is the shape parameter of the Moffat profile, here we fix it to be 2.2, according to our PSF modelling experience for KiDS images (e.g. \citealt{2018MNRAS.480.1057R}).} $\mu$ and $\sigma$ are the mean value and standard Deviation of a normal distribution}
\label{tab:parameters}
\end{deluxetable}

\begin{figure*}
    \centering
    \includegraphics[width=17cm]{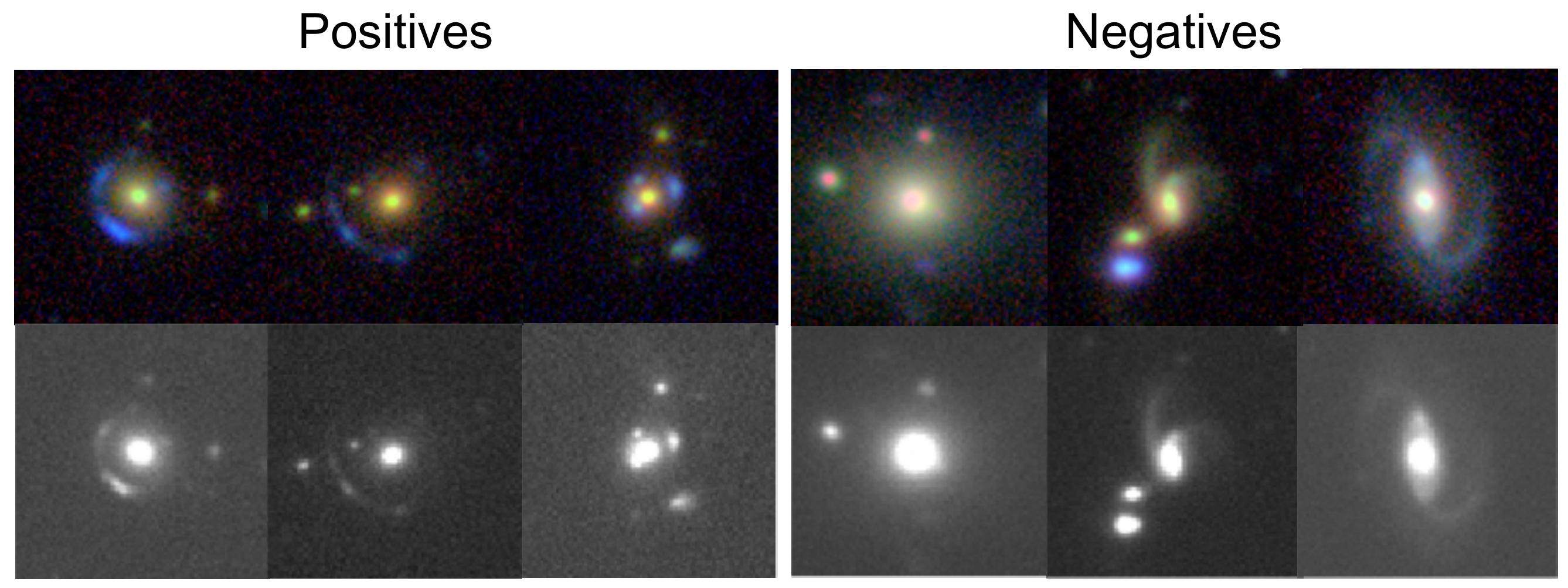}\vspace{3pt}
    \caption{Examples of the training sample. The three panels on the left show three simulated lenses (`positives') generated by adding mock arcs to real LRGs. The three right panels show three real galaxies used as `negatives'. The first row shows the $g,r,i$ color-composited images ($20"\times20"$) while the second row shows the $r$-band images ($20"\times20"$).}
    \label{fig:train_sample}
\end{figure*}

Once the simulated lensed images are obtained, we add these to 
randomly selected LRGs to generate real-like lenses. 
In this process, we require the ratio of the peak of the $r$-band surface brightness of the arcs $\alpha$ and the LRGs $\beta$ to satisfy the condition $\alpha/\beta \geq0.05$, or the local signal-to-noise ratio of the surface brightness peak of the arcs (3$\times$3 pixels with the peak as center ) to be ${\rm SNR}>5$. 
This is a less conservative choice than in L+20, where we used $0.02\leq \alpha/\beta \leq0.3$ {to }
avoid {the selection of extremely} faint simulated arcs that would be undetectable by human inspection.
However, this 
implies that the trained CNN is intrinsically less complete because it is unable to discover faint or even small separation arcs (see Sec \ref{sec:testing} for detail). The CNN is also
more accurate
in recognizing mid-/high-contrast arcs with color information. Indeed, we stress that larger completeness is generally obtained at the cost of a larger contamination, as most of the ``false positive'' are concentrated at faint magnitudes. 
Hence, with the new training sample, we aim at dramatically reducing the number of these contaminants by inhibiting the CNN to guess in presence of low-contrast, faint features. 

With these procedures, we finally build 43\,000 mock lenses as positives, together with the 43\,000 negatives, to train and test the CNNs. In Fig. \ref{fig:train_sample} we show a sample of positives and negatives from the training sample.

We decide to choose only LRGs to simulate the positives in the training and testing samples since they are generally the most massive systems and more likely to form SL events (e.g., SLACS, \citealt{2006ApJ...638..703B}; BELLS, \citealt{2012ApJ...744...41B}; LinKS, \citealt{2019MNRAS.484.3879P}).
This in turn 
reduces the number of ``false positives" on 
the final predictive sample on which the visual inspection is carried out. Nevertheless, in \cite{2020ApJ...899...30L} we have demonstrated that even without generalizing the training sample, the CNN can find high-quality lens candidates among the BGs sample. In fact, the majority of the HQ candidates found from the BGs sample are mainly identified thanks to the presence of sharp and bright {\it lensed} features, rather than the properties of the lens itself. Instead, by generalizing further the training sample, we would risk increasing the chance to add false negatives to the final list of HQ candidates.

Finally, during the training process, we optimize the CNNs with the Adam optimizer \citep{2014arXiv1412.6980K} by minimizing a `binary$\_$crossentropy'\footnote{https://keras.io/api/losses/probabilistic$\_$losses/$\#$binarycrossen\\tropy-class} loss. The learning rate is 0.0001.

\subsection{Testing the CNN}
\label{sec:testing}

\begin{figure}
    \centering
    \includegraphics[width=9cm]{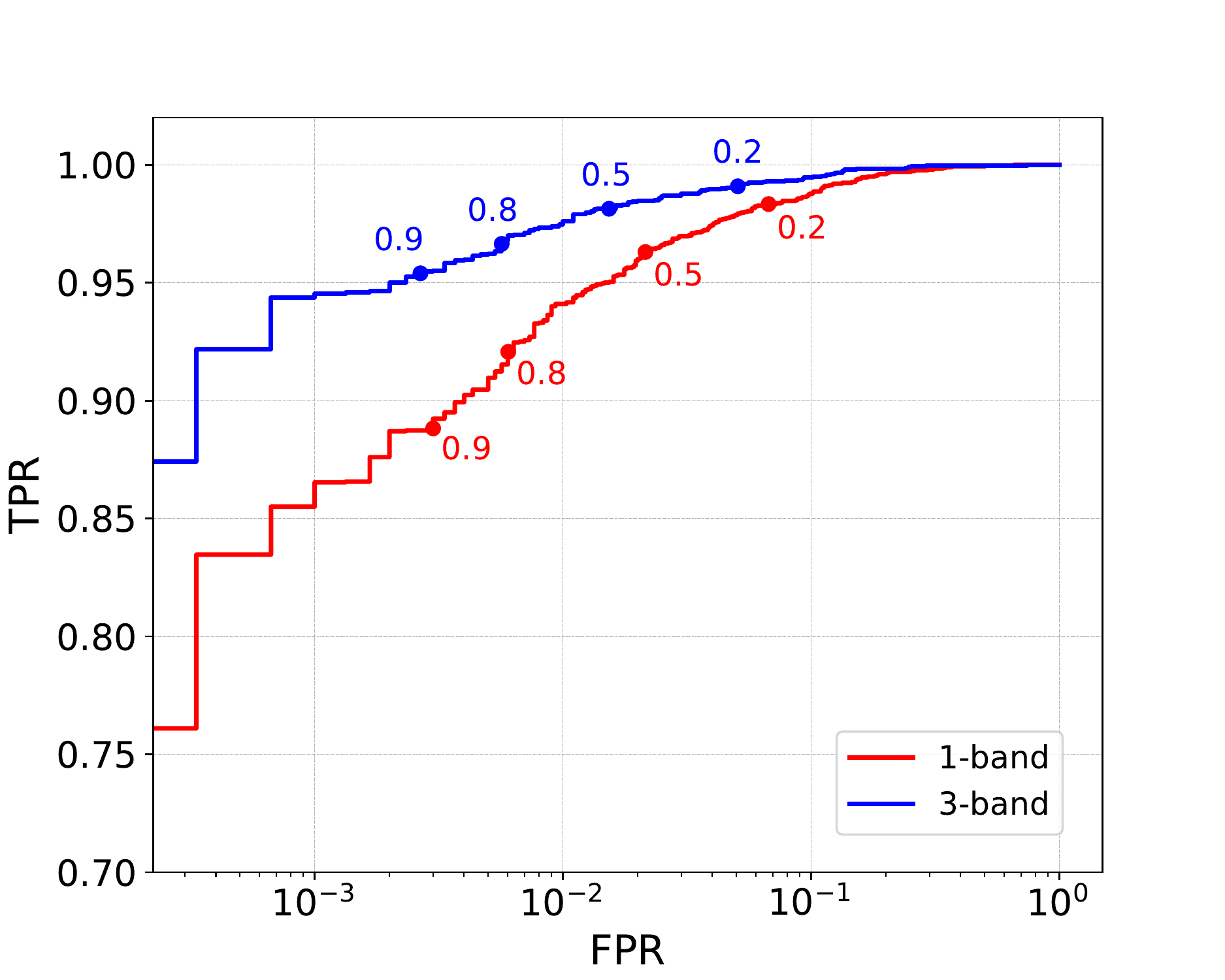}\vspace{3pt}
    \caption{The ROC curve for the $1$-band (red) and $3$-band (blue) CNNs classifiers based on {the } 3\,000 simulated lenses and 3\,000 galaxies {used } as testing sample. {On the curves, we also  }
    show the locations of 4 different values of probability threshold ($P_{\rm CNN}=0.2, 0.5, 0.8, 0.9$) used to calculate the FPR and TPR.}
    \label{fig:ROC_dist}
\end{figure}

{In each epoch of the training, we test the performance of the networks using the validation sample. This is needed to measure the accuracy reached by the CNN at every training step. When the accuracy reaches an asymptotic value, then the training process can be stopped.

After completing the training, we can assess the overall performance of the two CNNs on the test sample, and evaluate the ``false positive ratio'' (FPR) vs. ``true positive ratio'' (TPR) curves, where the TPR and FPR are defined as:
\begin{itemize}
\item {TPR}: The fraction of {\tt positives} that 
have also been identified as {\tt positives} by the classifier (i.e. objects on which the classifier works properly). 
\item {FPR}: The fraction of {\tt negatives} that have been wrongly classified as {\tt positives} by the classifier.
\end{itemize}
This FPR-TPR curve, also called the receiver operating characteristic (ROC) curve,} is useful to 
measure the contamination and accuracy of the classifiers. The ROC curves of the two CNNs are shown in Fig.~\ref{fig:ROC_dist}.
The introduction of the color information improves the {accuracy of the }
3-band over the 1-band CNN, as demonstrated by the higher number of true positives for the same false-positive ratios. This is {particularly true towards higher $P_{\rm CNN}$, where the $3$-band CNN reaches even higher true-positive rates for similar  false negative rates.  }

\vspace{5pt}

\begin{figure*}
    \centering
    \includegraphics[width=8.7cm]{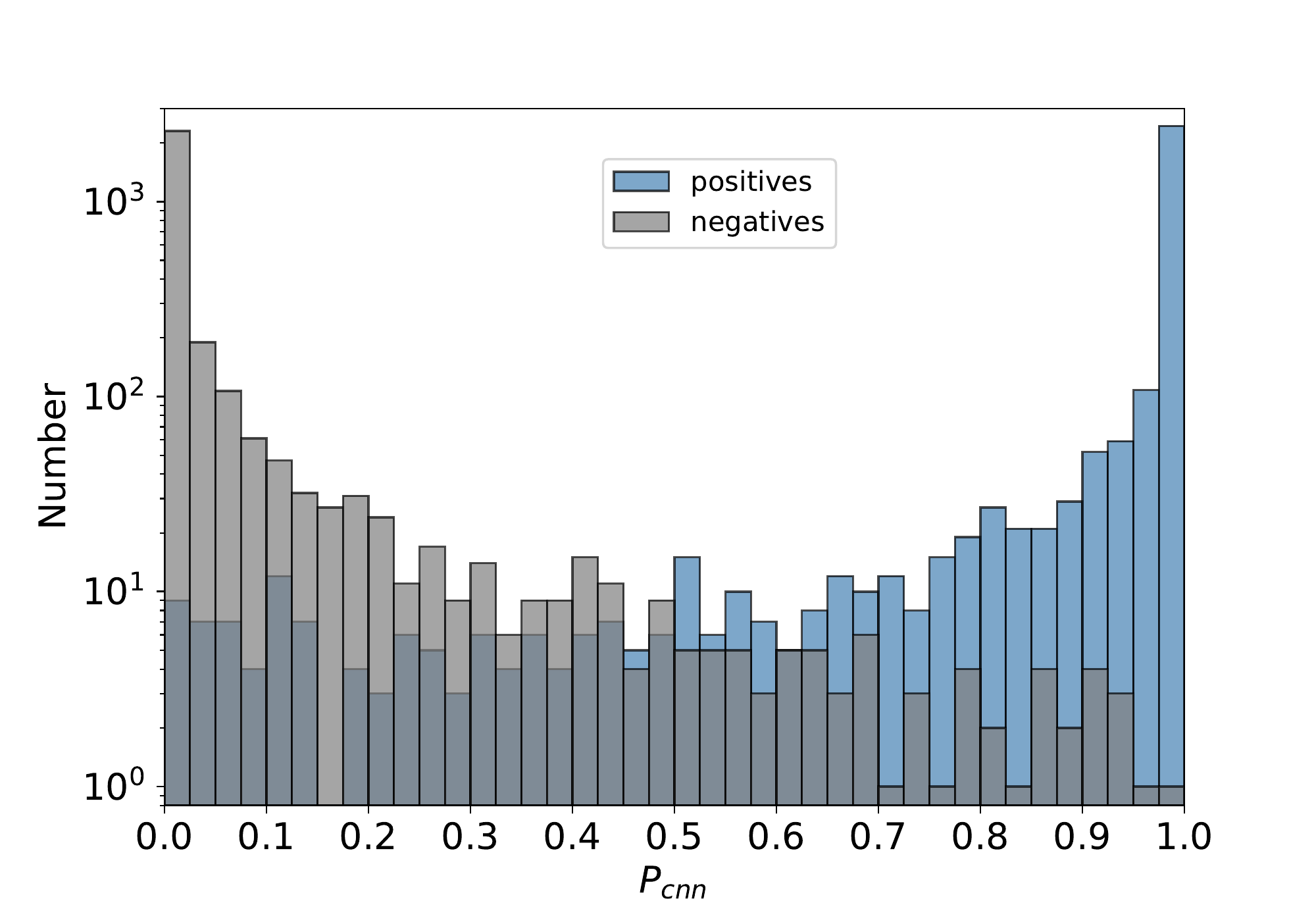}\vspace{3pt}
    \includegraphics[width=8.7cm]{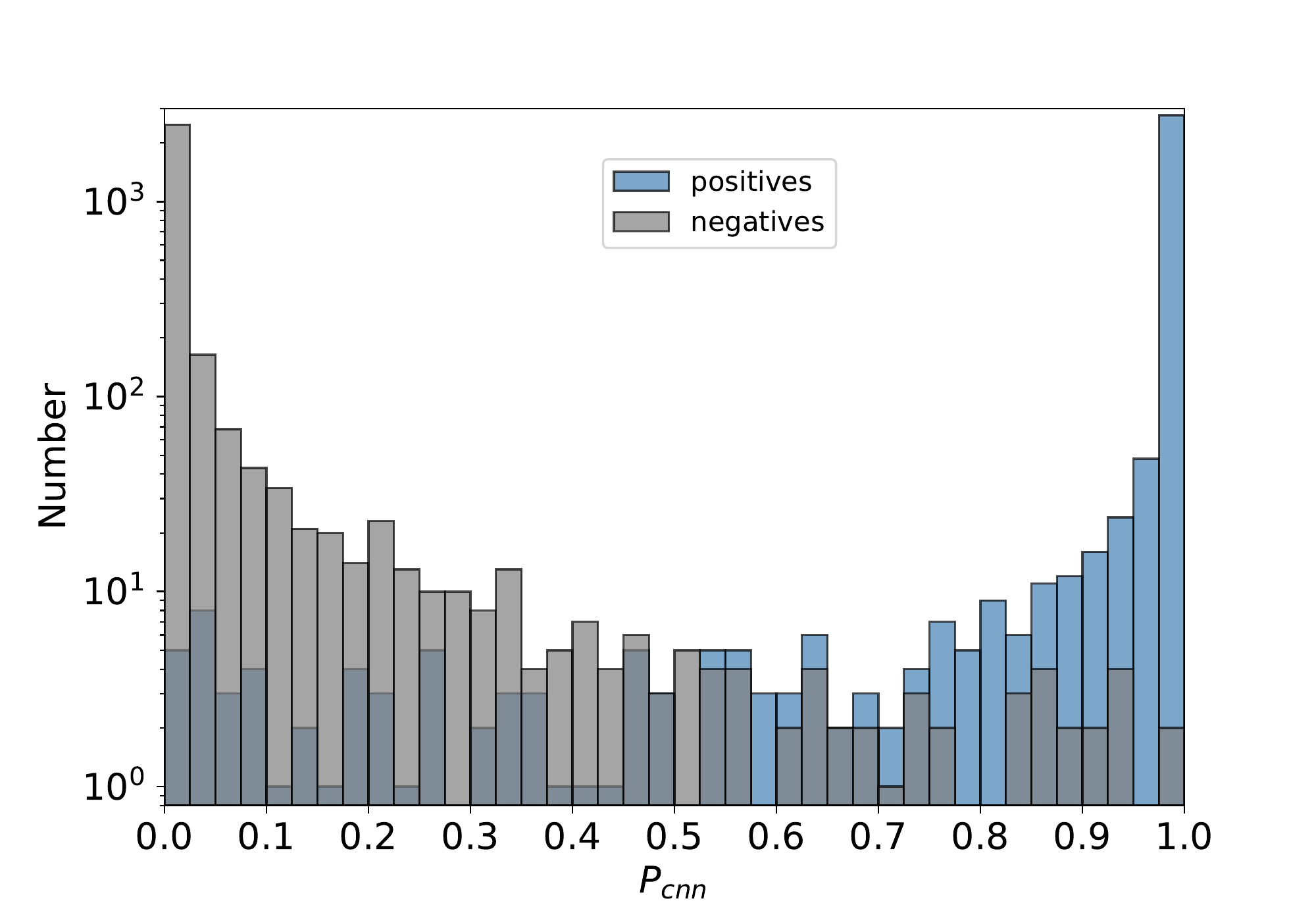}\vspace{3pt}
    \caption{The probability distribution of the testing sample for both 1-band (left) and $3$-band (right) CNNs. The blue histogram represents the probability distribution of the {\tt positives} while the grey histogram shows that of the {\tt negatives}.}
    \label{fig:test_prob_dist}
\end{figure*}

\begin{figure*}
    \centering
    \includegraphics[width=17cm,height=3.5cm]{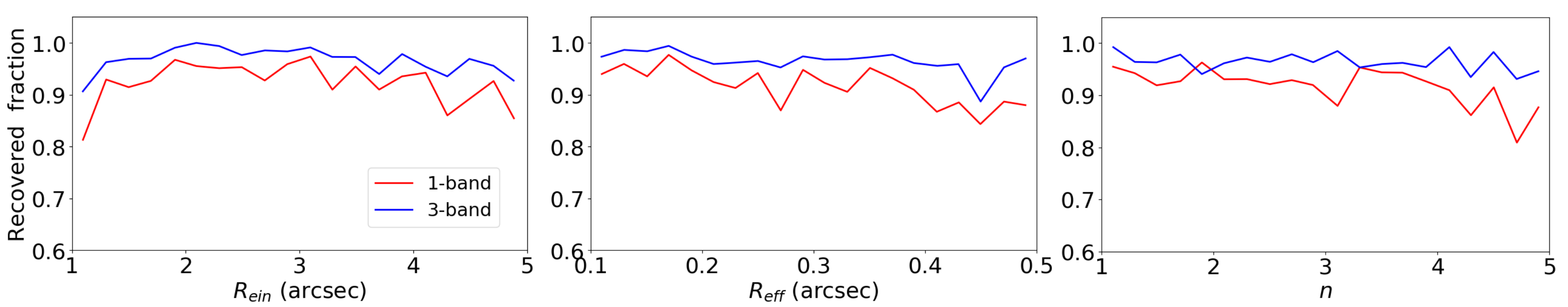}
    \setlength{\abovecaptionskip}{0.2cm}
    \caption{
    Fraction of the 3\,000 positives from the training sample that received a $P_{\rm CNN}>0.8$ in bins of Einstein radius (left panel), source effective radius (middle) and source S{\'e}rsic index (right). See text for details.
    }
    \label{fig:para_rec_frac}
\end{figure*}

In Fig.~\ref{fig:test_prob_dist} we show the $P_{\rm CNN}$ distribution for the 3\,000 positives (blue) and the 3\,000 negatives (grey) galaxies in the testing sample, {obtained for the 1-band CNN (left panel) and the 3-band CNN (right panel).}

\begin{figure}
    \centering
    \includegraphics[width=9cm]{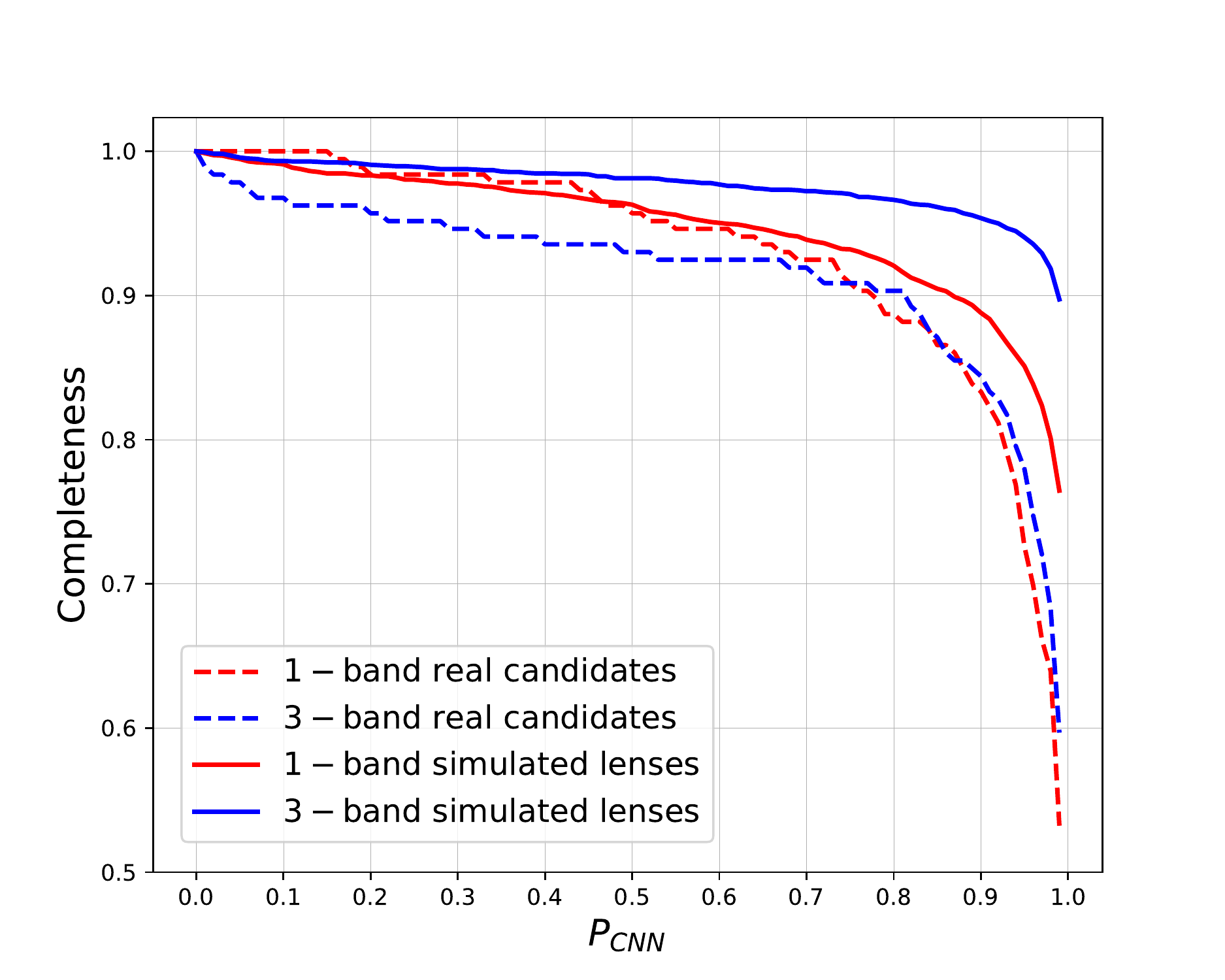}\vspace{3pt}
    \caption{The completeness of the $1$-band (red) and 3-band (blue) classifier at different probabilities based on 3\,000 simulated lenses (solid lines) and 179 real high quality lenses candidates (dash lines). The real lens candidates are collected from P+19 and L+20.}
    \label{fig:completeness_real_data}
  \vspace{0.3cm}
\end{figure}

In the ideal case, the positives should all be clustered around $P_{\rm CNN}=1$, while all negatives should peak at $P_{\rm CNN}=0$.
The closer to the ideal case a CNN classifies these two categories, 
the better it performs. {In Fig.~\ref{fig:test_prob_dist},} 
both CNNs clearly identify the easiest cases as there is a peak of positives at $P_{\rm CNN}=1$ and a peak of negatives at $P_{\rm CNN}=0$. However, the 3-band CNN shows a steeper decline of the distributions of both categories and has a small number of wrongly classified objects. This means that the 3-band CNN has a sharper ability to distinguish positives and negatives.
As a consequence of that, the 3-band CNN concentrates most of the positives at higher $P_{\rm cnn}$ ($>0.5$), where 
there are only a few negatives, which
possibly will end up to be false positives {in the final catalog}. This explains the better performance seen on the ROC curve, where a larger TPR is found for the 3-band CNN, especially at higher $P_{\rm CNN}$ values.   
Even if the 1-band CNN performs almost equally well in terms of false positives (i.e. few negatives with rather high probability), it spreads {the} true positives over a broader range of low $P_{\rm CNN}$. This {affects }
the 
completeness in the recovery of the ``true lenses'' {since some of them would not pass the probability threshold.}
{The completeness, defined as the fraction of recovered$/$simulated lenses, is shown in Fig.~\ref{fig:completeness_real_data} as}
a function of the output probability $P_{\rm CNN}$ of the two classifiers. As expected, the 3-band classifier performs systematically better than the 1-band classifier, and reaches 90\% completeness for a $P_{\rm CNN}=0.98$, while the 1-band CNN reaches the same completeness for $P_{\rm CNN}>0.85$.
This shows that the colors of the lensing features add  relevant information to correctly classify the positives 
and separate them from the negatives.

In order to know how the input parameters affect the completeness of the classifier, in Figure \ref{fig:para_rec_frac} we plot the recovered fraction ($f1$ for 1-band classifier and $f3$ for 3-band classifier) of the 3\,000 testing positives in different bins of Einstein radius, source effective radius, and S\'ersic index, respectively. From the figure, we can identify two main features. First, lenses with very small Einstein radii ($R_{\rm ein}\lsim1.5''$) have lower recovered fraction. This is because small separation arcs are more likely to be hidden behind the light of the deflectors, and it is difficult for the classifier to deblend them from the foreground galaxies. 
However, 
unexpectedly, 
after both $f1$ and $f3$ reach a peak around $R_{\rm ein}=2''$, they start to decrease for larger Einstein radii ($R_{\rm ein}>3''$). This effect is rather weak, but systematic, especially for the 1-band CNN, and can be a combination of confusion effect and poor sampling of large $R_{\rm ein}$ in the training sample. A similar trend was also found in (\citealt{2019MNRAS.482..807P}). The confusion effect comes from the fact that for larger separation there is a higher 
chance for the presence of companion or contaminant galaxies, even in the training sample. This makes the classification more uncertain 
and the $P_{\rm CNN}$ lower. The poor $R_{\rm ein}$ sampling is a choice due to simulation realism and reflects the fact that these lenses are rarer (see \citealt{2015ApJ...811...20C}). We note that high separation lenses tend to receive lower scores also from human inspectors, for the same reasons. 
Second, $f3$, which does not correlate with the source effective radius and the S\'ersic index, is generally higher than $f1$, which instead shows a decreasing trend with both source parameters. This demonstrates that source images that are generally more compact and sharp (like the ones obtained by compact galaxies or disks) are easier to be identified by the 1-band CNN, which is less performing with more diffuse arcs. Here the 3-band CNN compensates with the color information and hence becomes more efficient.

Overall, Figures \ref{fig:ROC_dist}, \ref{fig:test_prob_dist} and \ref{fig:para_rec_frac} show that the 3-band CNN really benefits from the multi-band information, despite the fact that the $g$ and $i$-bands have seeing generally worse than $r$-band,
hence reducing the sharpness of the lensing features to be identified in the images. 
Likely, the color information compensates of this blurring effect, producing generally higher $P_{\rm CNN}$.

\begin{figure*}
    \centering
    \includegraphics[width=17cm]{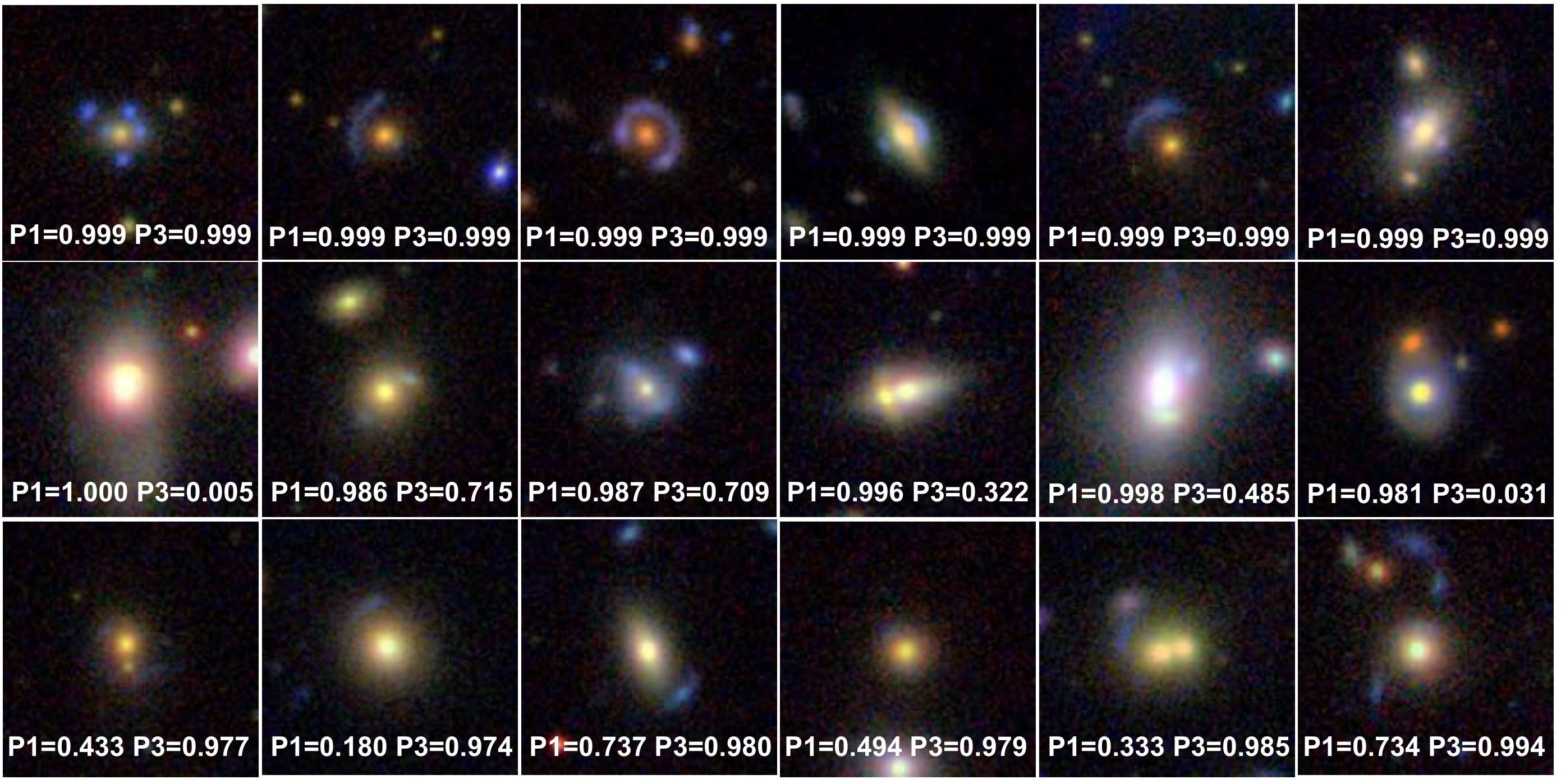}\vspace{3pt}
    \caption{HQ candidates from P+19 and L+20 that received different predictions from the $1$-band classifier and 3-band classifier. Candidates in the first row obtained high probabilities from both the $1$-band and 3-band classifier ($P_1>0.99$ and $P_3>0.99$). {In this group of objects }
    there are almost no 'false positives'. Candidates in the second row obtained higher probability from the $1$-band classifier but a lower one from the 3-band classifier ($P_1>0.95$ and $P_3<0.8$). {Candidates }
    shown in the third row obtained lower probability for the 1-band classifier but a higher probability for 3-band classifier ($P_1<0.8$ and $P_3>0.95$. The stamps (2''x2'') are obtained by combining $g$, $r$, and $i$ KiDS images.}
    \label{fig:P1_P3_discrepancy}
\vspace{0.6cm}
\end{figure*}

However, this over-performance of the 3-band CNN can be the consequence of the usage of rather ideal strong lensing configurations as the test sample, namely simulated arcs/multiple images (see Fig.~\ref{fig:train_sample}). Hence, we should use caution by expecting a similar performance in real cases. This is a general problem of the true performance of CNNs trained on simulated datasets
but then applied to real lenses.  
As a sanity check on the reliability of such performance, we apply
the two CNNs to 179 high-quality lens candidates from \cite{2019MNRAS.484.3879P} and \cite{2020ApJ...899...30L}. 
In Fig.~\ref{fig:completeness_real_data}, we overplot the completeness obtained for the ``real candidates'' 
to the one obtained for the test {(simulated)} sample.
Overall, the two CNNs perform in a much more comparable way than they
do on the test sample.  
The 3-bands CNN has a {slightly }higher completeness at $P_{\rm CNN}>0.75$, while 
the 1-band CNN does slightly better at lower probabilities. 

However, the degradation of the performance of the 3-band CNN, from the mock lenses to the real ones, seems particularly worrisome, especially because this is not mirrored by the same effect on the 1-band. 
This might suggest a problem with the mock colors. 
To check that, in Fig.~\ref{fig:P1_P3_discrepancy} we show some examples of ``real candidates'' with their probabilities ($P_1$ for 1-band CNNs and $P_3$ for 3-band CNN). 
{In particular, in the first row, we show images of candidates that received very high scores from both CNNs. They all show clear {blue and bright} lensing features.}
In the second row, we {show instead} candidates for which $P_3\lsim0.5$ and $P_1>0.98$. They also have
clear arcs
but with either yellow/red colors or shallower bluer images, which return a low-$P_3$ because our training sample is tailored on bright bluer sources. 
These kinds of
candidates {which will be discarded since they received a low grade,} can potentially reduce dramatically the completeness of the 3-band candidates when applied to real galaxies. 
This is a weakness of {our }
multi-band design
that we will study in future work.

Finally, in the third row of the same figure,
we show the candidates with high $P_3$ but low $P_1$, which instead demonstrated why higher completeness of the 3-band CNN is reached at higher $P_{\rm CNN}$. In this case the 3-band CNN tends to select also arcs or multiple images that the 1-band CNN ranks lower. This is because the 3-band CNN has been trained to search for sharpness features in the bluer bands, due to the varying FWHM in Table~ \ref{tab:parameters}, as discussed in Section \ref{sec:training_data}. {Instead, the }
1-band CNN returns a low $P_{\rm CNN}$ because the lensing features are fainter and less defined in the $r$-band. 
Overall, Fig.~\ref{fig:P1_P3_discrepancy} demonstrates that the two CNNs are complementary to each other, as there are good candidates that received very high probability in one case but very low probability in the other, and vice-versa.
This ``complementarity'' mainly resides in the different features/information the two CNN are based on. 

\begin{figure}
    \centering
     \includegraphics[width=9.5cm]{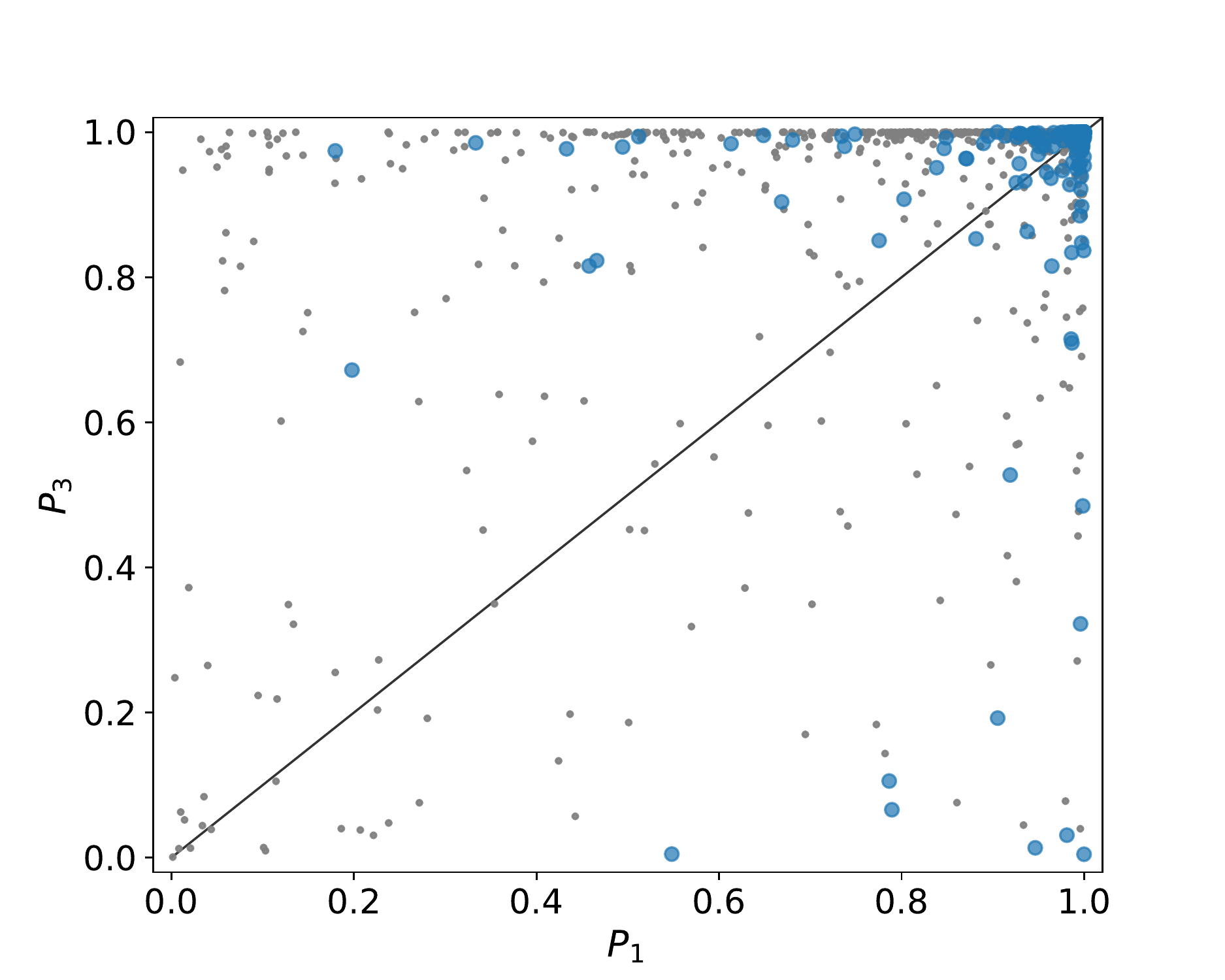}\vspace{3pt}
    \caption{{Probability of the real lens candidates (blue) and 3000 mock lenses (grey) for the 1-band versus the probability for the 3-band CNN.}
    The two probability look 
    uncorrelated, suggesting a large complementarity between the two methods if applied to ground based observations. See discussion in the text. }
    \label{fig:compare_P1_P3}
\end{figure}

To further quantify that,
in Fig. \ref{fig:compare_P1_P3} we compare the probability distribution of the ``real'' lenses (blue dots) obtained from the $1$-band CNN versus the ones obtained from the $3$-band CNN.
Encouragingly, most of the 60\% of the lens candidates are located in the high $P_{\rm CNN}$ ($>0.95$) corner, meaning that the SGL features have both clear morphology and color contrast, with the lensed images being bluer than the lens. 

For $P_1,~P_3<0.9$, the scatter between the probabilities of the two CNNs becomes large. More importantly, the points are not distributed around
the 1-to-1 relation, which would imply a 
consensus between the two classifiers. 
Instead, 
there is a number of lenses that are highly ranked by one CNN but received a very low score from the other one.
In particular, if we use a threshold of $P_{\rm CNN}=0.8$ for both CNNs, 21 objects will be discarded from the $1$-band CNN and 18 will be discarded from the 3-band CNN. But only 6 of them would be discarded by both.
This means that there are {33} HQ candidates (i.e. 18\% of the total sample) which represent a ``complement'' sample that would be selected by only one CNN\footnote{We remark that this ``real sample'' has been selected by different CNNs: a pure 1-band CNN (L+20) and a mix of 1-band and 3-band CNNs (P+19), and visually inspected. Hence, the lack of predicted objects along the 1-to-1 relation is a consequence of the human inspection that tend to ``select'' candidates either toward higher $P_1$ or high $P_3$.}.    
In Fig.~\ref{fig:compare_P1_P3} we also show the same comparison for the test sample (grey dots) of LRGs, which corresponds to the majority of systems the ``real sample'' has been selected from. Here we clearly see the only small correlation between the output $P_{\rm CNN}$ of the two classifiers, confirming the strong complementarity of the two approaches. However, a large number of points are visible with $P_3>0.95$ at
almost all $P_1$. We also see a fair number of $P_1>0.95$ at $P_3<0.8$. If we choose to select candidates which are above a conservative threshold on both classifiers, we would lose a lot of these two tails, which are mirrored by the distribution of the ``real lenses''. 
For instance, for $P_1>0.8$ and $P_3>0.8$, we would select only 2734 positives,
i.e. 46\% of the whole 6000 testing sample (3000 positives plus 3000 negatives).  On the other hand, we can take advantage of the segregation toward the tails in high $P_1$ and $P_3$ to optimize the candidate selection and the overall completeness of the two CNNs together.
{In conclusion, the best strategy to take full advantage of their complementarity is to select LRG candidates above a given threshold for one {\it or} the other CNN. }In 
Section \ref{subsec:predictions} we will introduce a more conservative approach to the BG sample.

\section{Applying the classifiers to KiDS-iDR5}
\label{sec:prediction_DR5}
KiDS-iDR5
includes a total of 1347 tiles, of these 1006 have been publicly released as data release 4 (DR4) 
and searched for strong lenses
in P+19 and Li+20. In this paper, we therefore
analyze the remaining 341. 

\subsection{Selection of the predictive samples}
\label{subsec:predictive}
We define two
``predictive samples'', 
in which we expect to find SGL events, using the method described in \cite{2017MNRAS.472.1129P, 2019MNRAS.484.3879P} and Li+20. 
We first define the larger sample of BGs as those objects with
KiDS star–galaxy separation parameter $2DPHOT=0$ (\citealt{2008PASP..120..681L, 2015A&A...582A..62D}) and $r-$band Kron-like magnitude (\citealt{1996A&AS..117..393B}) $mag\_auto \leq 21$.

Then, we define the LRG sample, that is selected from the BGs by setting two further criteria: 1) $r_{\rm auto} \leq 20$, 2) the colors satisfy a slightly adapted LRG color–magnitude selection (see also in \citealt{2001AJ....122.2267E}, {P+19 and Li+20}):
\begin{equation}
\begin{split}
r_{\rm auto}<14+c_{\rm par}/0.3,\\
|c_{\rm perp}|<0.2,
\end{split}
\label{eq:cuts}
\end{equation}
where
\begin{equation}
\begin{split}
c_{\rm perp}=(r-i)-(g-r)/4.0-0.18,\\
c_{\rm par}=0.7(g-r)+1.2[(r-i)-0.18].
\end{split}
\label{eq:colours}
\end{equation}
\rui{Different galaxies at different redshifts ($z<0.4$) form a nearly 1-dimensional locus in $g-r$ and $r-i$ space (see detail in \citealt{2001AJ....122.2267E}). Here, $c_{\rm perp}$ and $c_{\rm par}$ measure the position along and across this locus in a rotated coordinate system in color-space.

                                                                                                                                                                                                                                                                                                     With these criteria, we identified 72\,010 LRGs and 1\,432\,348 BGs for the 341 tiles.}

\subsection{1-band and 3-band CNN predictions}
\label{subsec:predictions}
{We first apply the CNNs described in Section \ref{sec:new_classifier} to the LRG sample. In order to achieve $\gsim 90\%$ completeness, we set the threshold probability to be $P_{\rm CNN}>0.8$ for both CNNs (see Fig.~\ref{fig:completeness_real_data}). 
We retrieve candidates passing one {\it or} the other classifier, to take advantage of the complementarity between the two CNNs, as demonstrated in Section~\ref{sec:testing}.}
The number of candidates 
is 1213 (1.67\% of the LRGs) and 1299 (1.80\% of the LRGs) from the 1-band classifier and 3-band classifier, respectively, with an overlap of 442 {systems} ($\sim30\%$).
This overlap is smaller than what was found for the ``test sample'' in Section~\ref{sec:testing}. This is because in the testing phase we could exclude contaminants, which we cannot do on real data. As shown in Fig.~\ref{fig:train_sample}, we expect some spurious candidates also at the high-end of the $P_{\rm CNN}$ distribution.
The final list of candidates, after removing duplicates, comprises 2070 automatically selected galaxies.

We also apply the two classifiers to the BG sample. Considering that the CNNs have been trained only on LRGs.
This implies higher contamination for BGs with respect to LRGs, at any given predicted $P_{\rm CNN}$. 
Hence, with the aim of maximizing the number of HQ candidates passing the threshold, while minimizing the number of possible contaminants, we use a higher threshold, selecting only candidates with $P_1>0.9$ {\it and} $P_3>0.9$. Such a strategy was proved to be successful in L+20, where we found very HQ candidates within this galaxy selection. This returns 3740 new lens candidates from the BG sample.

\subsection{Visual inspection and selection of HQ candidates}
\label{subsec:visual}

\rui{After removing all overlaps, we finally obtain 5810 ($1213+1299-442 LRGs+3740 BGs$) candidates.
These candidates are initially inspected by one observer\footnote{RL, leading author of this paper.}, to further clean the sample from secure non-lenses.}
The number of candidates that have survived this process is 487, among which 192 from the LRG sample and 295 from the BG sample. 
We define this as the ``good quality'' sample.

Seven inspectors\footnote{These are RL, NRN, CS, CT, LL, WS, GC.} visually grade the ``good quality'' sample, according to an $ABCD$ quality letter scheme where $A$ is a sure lens, $B$ is maybe a lens, $C$ is maybe not a lens and $D$ is not a lens{. We subsequently associate to each letter a } mark of 10, 7, 3, 0 respectively, to convert the quality flags into a score (see L+20). 
Furthermore, to reduce the impact of biased judgment from any of the inspectors on the final visual inspection score,
we decided to remove the highest and the lowest scores {for each candidate,} and compute the average scores, $s_{\rm ave}$, from the remaining 5 inspectors. 

In Fig.~\ref{fig:high_quality_candidates} we show the 97 candidates that have received $s_{\rm ave}\geq6$, which we define as HQ lens candidates. 
In Table~\ref{tb:lens_candidates} we list the ``KiDS$\_$ID", coordinates, $r$-band magnitudes, probabilities from CNNs, and scores from human inspection

Among these, we identify some interesting systems. J234804.98$-$302855.45 is a new Einstein cross, that, based on the color and size of the quadruple images, seems to be a lensed blue nugget system like the ones we have found in \cite{2020ApJ...904L..31N}.
J112610.75+033847.89 is a large arc around a close pair of galaxies.

The CNN has also identified strong lenses in galaxy clusters/groups, e.g., J020706.67$-$272644.71, J124159.31$-$021756.09, J014320.63$-$271746.99, and J021227.44$-$284258.00. These show clearly visible arcs around a few bright galaxies in a crowded field, suggesting a cluster environment. These systems are just a demonstration of the variety of deflectors and sources accessible with the new data, and the science one can pursue 
with extended
spectroscopic follow-ups. 
Indeed, 
the collection of large numbers of HQ candidates  provides
statistical samples for DM studies in the lenses in isolation and in denser environments, as well as
interesting targets to address galaxy formation issues at high redshift with lensing magnification effect. Spectroscopic confirmation remains the only missing information for these targets. 

\section{Discussion}
In this section, we discuss in more detail the two major outcomes from this work: the discovery of new HQ lens candidates, which we make publicly available; and the comparison between the performance of the $1-$band versus the $3-$band classifiers on ground-based observations.

\subsection{The first KiDS catalog of HQ lens candidates}

Putting together the new HQ candidates presented in this work with the ones previously found in KiDS (\citealt{2017MNRAS.472.1129P, 2019MNRAS.484.3879P}, L+20), the number of HQ candidates in the survey has reached 268. This first catalog of strong lens candidates in the final KiDS footprint is made available at this link \href {https://kids.strw.leidenuniv.nl/DR4/hqlenses.php}{https://kids.strw.leidenuniv.nl/DR4/hqlenses.php}.

These HQ definition candidates have a very high probability of being confirmed, as demonstrated by the follow-up observations performed so far, that generally return a $\gsim$70\% confirmation rate (see e.g., \citealt{2019MNRAS.483.3888S}, \citealt{2020MNRAS.494.3491L}, \citealt{2020MNRAS.494.1308N}, \citealt{2020ApJ...904L..31N}). 
The ``good candidates'', instead, generally have a confirmation rate of the order of 40\% or lower, making them suitable as ``filler'' targets for large sky spectroscopic surveys.

Our new KiDS findings are meant to contribute to the ongoing effort 
of the scientific community to collect the best candidates needed for the planning of large-scale spectroscopic follow-up, e.g., with 4-metre Multi-Object Spectroscopic Telescope ( 4MOST, \citealt{2019Msngr.175....3D}).

Indeed, the majority of currently available candidates recently found in multi-band wide-sky surveys (e.g., \citealt{2020A&A...644A.163C, 2019ApJS..243...17J, 2021ApJ...909...27H}), is yet to be confirmed, thereby limiting the scientific outcome of these lens samples. 
Instead, large samples of confirmed SGL events are utterly required.
First, it would make process on a variety of science cases, from astrophysics to cosmology (see Section \ref{sec:intro}). Second, a large database of confirmed SGL systems is necessary to optimize the training of CNNs for future large sky surveys. In future space missions (e.g. Euclid or CSST) and ground-based observations (e.g. Rubin/LSST) we will face the challenge of selecting up to $10^5$ lenses out of tens of million of candidates, and for this we will need highly accurate classifiers to make both the search and the spectroscopic follow-up feasible,efficient and accurate. 

In this context KiDS candidates
offer unique advantages. 
Compared with existing lens samples imaged by the Hubble Space Telescope (e.g., SLACS,  \citealt{2008ApJ...682..964B}; BELLS, \citealt{2012ApJ...744...41B}, BELLS GALLERY, \citealt{2016ApJ...833..264S}), KiDS, combined with the twin NIR VIKING survey \citep{2013Msngr.154...32E}, provides us photometry images in 9 bands ($u, g, r, i, Z, Y, J, H, Ks$), which can be used for accurate stellar mass establishment of both deflectors and sources (see e.g. \citealt{2020ApJ...904L..31N}). 
This is a crucial ingredient to derive unbiased DM estimates of the lenses and
characterize the physics of the high redshift galaxies in the background. 

\begin{figure}
    \hspace{-0.5cm}
    \includegraphics[width=8.7cm]{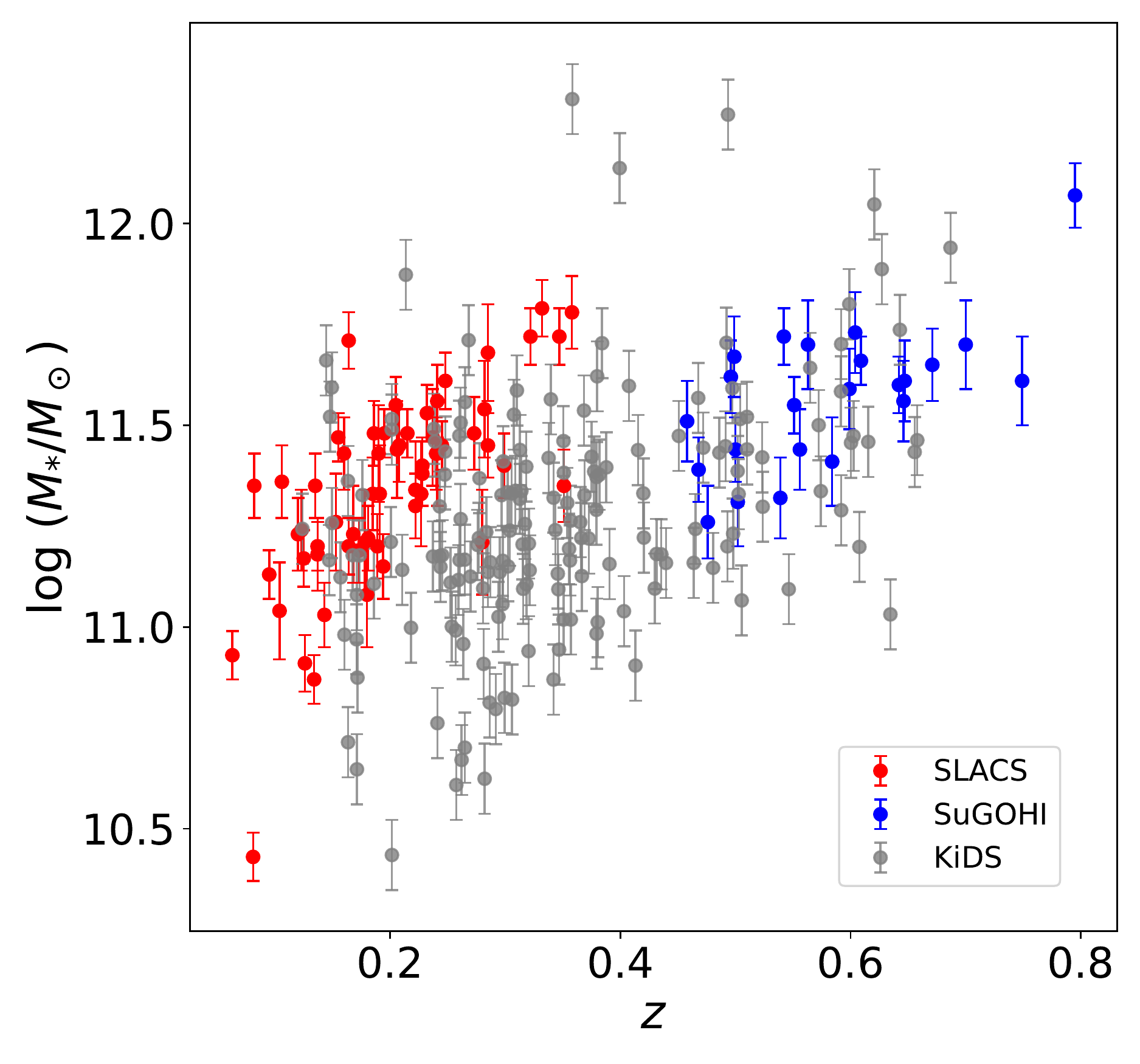}\vspace{3pt}
    \caption{Stellar mass versus redshift {for KiDS HQ lens candidates (grey dots), and }
    confirmed lenses from SLACS (red symbols) and the SuGOHI (blue symbols).
    For KiDS candidates we use photometric redshifts determined by machine learning method 
    \citep{2021inas.book..245A}  and stellar mass from 9-band photometry SED fitting.} 
    \label{fig:z_mass}
\end{figure}

KiDS lenses have a wider stellar mass coverage, which can be used, in particular, to push the study of the DM content of galaxies toward the sub-$L^*$ regime, where a transition of the DM properties has been observed (e.g., \citealt{2019MNRAS.489.5483T, 2013MNRAS.432.1709C}). This can be seen in Fig.~\ref{fig:z_mass}, where we plot the stellar mass versus photometric redshift (photo-$z$) of 174\footnote{5 candidates are removed because of the lack of photometry in the NIR that prevented the photo-$z$ inferences.} HQ lens candidates used to test the CNNs in Section~\ref{sec:testing} (grey dots).
The photo-$z$ of KiDS lens candidates are derived from the 9-band photometry using {\tt METAPHOR}{,} 
a machine learning tool for photometric redshifs (\citealt{2017IAUS..325..197A,2021inas.book..245A}). Stellar masses are obtained
by spectral energy distribution (SED) fitting of the 9-band photometry  using {\tt Lephare} (\citealt{2006A&A...457..841I}). For the latter, we adopt the \citet{2003MNRAS.344.1000B} stellar population synthesis models and assume a \citet{2003ApJ...586L.133C} stellar Initial Mass Function. As uncertainties on the stellar masses, we consider a 0.2 dex mean error, which reasonably accounts for modeling methods, and photometry and redshift uncertainties (see \citealt{2019A&A...632A..34W} for details).

In the same figure, in red and blue we overplot confirmed gravitational lenses from the Sloan Lens ACS Survey (SLACS, \citealt{2006ApJ...638..703B}) and the Survey of Gravitationally-lensed Objects in HSC Imaging (SuGOHI, \citealt{2019A&A...630A..71S}), respectively. SLACS objects cover a lower redshift range and have a mass distribution which is more biased toward high stellar masses, whilst \rui{the SuGOHI lenses} are restricted to $z\ge0.4$ and do not cover the lowest mass bins. Compared to both datasets, KiDS shows a more extended distribution both in redshift and stellar masses. 

Moreover, the 9-band photometry offers a unique opportunity
to exploit the use of accurate photo-$z$s of both lenses and sources (Li et al. in preparation). These can be used to have a first photometric {solid indication of the true lensing nature of the system, }
hence providing an ultra-refined sample for deeper spectroscopic follow-up.
Finally, (unbiased) photo-$z$s can be used in the ray-tracing models to derive (unbiased) mass estimates with low but reasonable accuracy ($\sim20-30$\%), which might be mitigated if large statistical samples are gathered. 

Hence, the KiDS sample offers the possibility to enhance both detection techniques and mass modeling techniques.
The optical+NIR multi-band photometry will also be available in combined datasets from future facilities (e.g. Rubin/LSST or CSST in the optical + {\rm Euclid} in NIR).

\subsection{1-band vs. 3-band CNN performance}
\label{subsec:discussion_of_performance}

\begin{figure*}
    \centering
    \includegraphics[width=8.5cm]{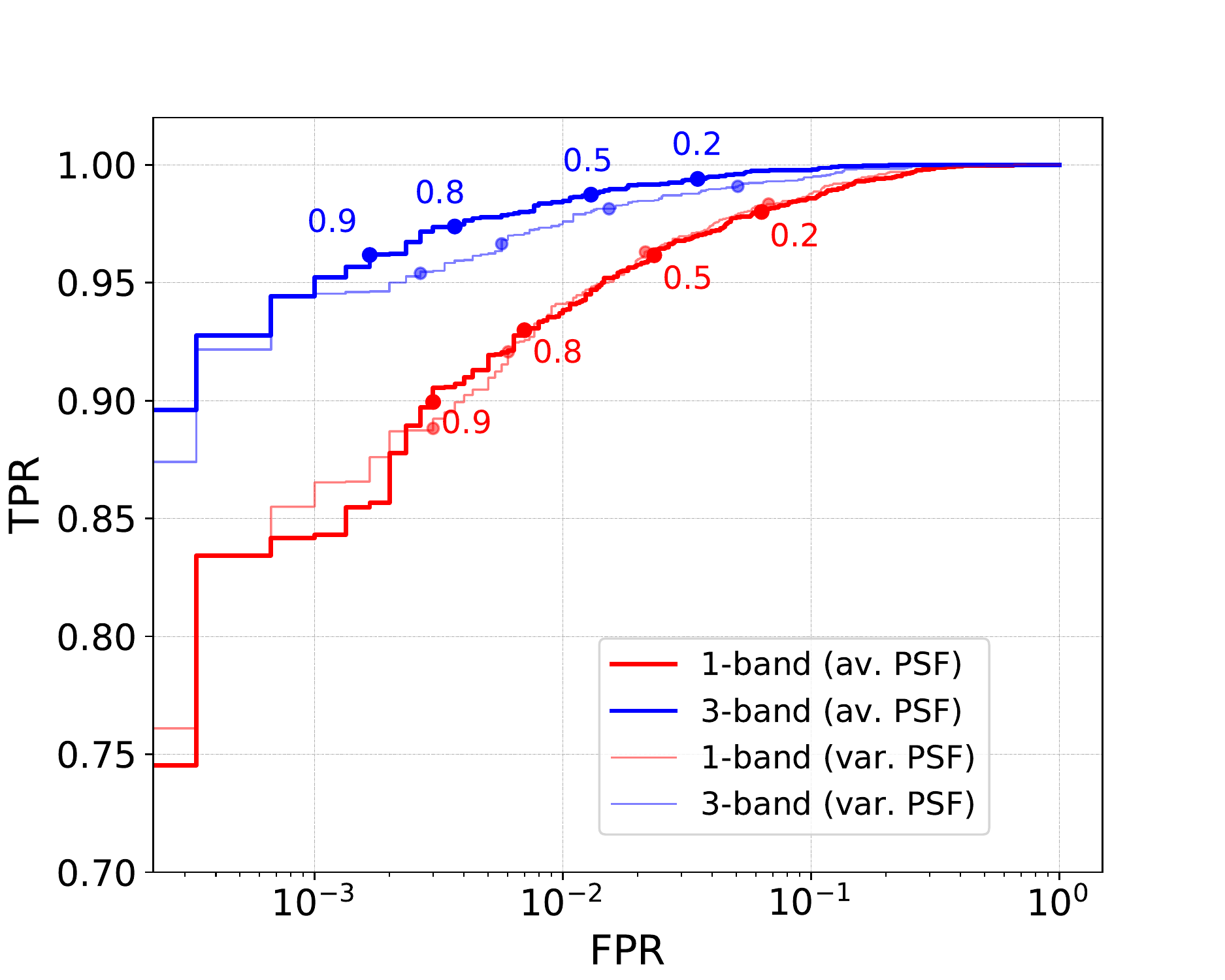}\vspace{3pt}
    \includegraphics[width=8.5cm]{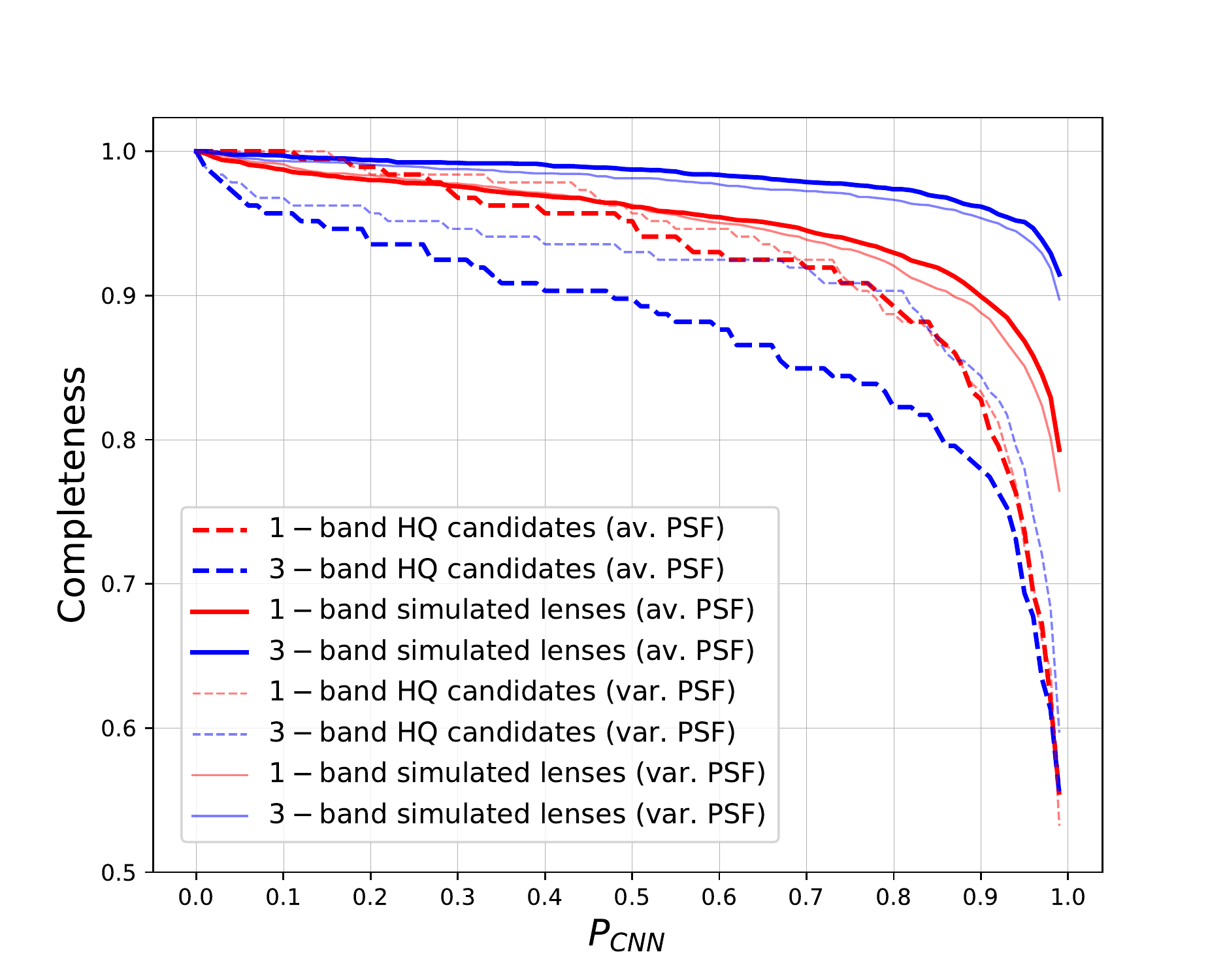}\vspace{3pt}
    \caption{{ROC curve (left) and completeness (right) of the CNNs for average PSFs (thick lines) and variable PSFs (thin lines, equivalent to Figs~\ref{fig:ROC_dist} and \ref{fig:completeness_real_data}, respectively). In both panels, the red lines show the results obtained for the $1$-band CNN and the blue lines these obtained for the $3$-band CNN. The completeness on HQ candidates for the $3$-band CNN improves using a variable PSF. All the other curves show only minimal changes.}}
    \vspace{22pt}
    \label{fig:completeness_average_psf}
\end{figure*}

An important result of this work is the comparison of the performance of the CNN {that uses only } 
high-quality $r$-band images (seeing $\lsim0.8''$ in KiDS) and the one using color information. 
In Section~\ref{sec:testing} we have shown that the two CNNs are complementary because they recognise lenses based on different features. The 
1-band case focuses on the morphology of the strong lensing features, the 3-band classifier on the color contrast between the red lens and the bluer arcs or multiple images. Ideally, one expects that the color information should improve the detection of {lensing }
features.

However, 
although colors do improve the performances on simulated data, there is no corresponding increase incompleteness in the real data (see Fig.~\ref{fig:completeness_real_data}).
Here we discuss in more detail two possible reasons behind this result. 

The first reason is empirical 
and related to the color library used to create multi-band images. 
This has been created from LSST simulations, which could reproduce a different color distribution {with respect to that covered by KiDS galaxies, }
thus biasing the efficiency of the 3-band CNN when applied to KiDS data. 
In order to explore this possible effect, we have tested a different color library, based on COSMOS models in {\tt Lephare} (\citealt{2006A&A...457..841I}),
as previously done to simulate arc colors in {P+19}.
From the COSMOS library, we selected models with galaxy types later than S0 and calculated the observed-frame magnitudes in the KiDS $g, r, i$ bands for redshifts from 1 to 3 in steps of 0.02. From these, we produced a new color library to create the lensed images and repeated the training, validation and testing of the 3-band CNN. We have finally applied the new ``COSMOS-color'' CNN 
to both the simulated data and the real data and found no significant difference in terms of $P_{\rm CNN}$ with respect to the results obtained with the LSST library. 
However, we need to remark that the majority of the arcs simulated from these two color libraries are blue, which 
means that the 3-band CNN is expected to have a limited capability to identify lenses with ``non-blue'' arcs. 
In Section~\ref{sec:testing} we have anticipated this lower sensitivity to red or yellow arcs, compared to the 1-band classifier.
In the future, we plan to extend the color library also to redder colors, to improve the completeness of the 3-band CNNs on real data. 

The second reason behind the degradation of the 3-band CNN performance on the real data
is methodological
related to the matching of the PSFs adopted for the simulated arcs and the actual PSF of the hosts. 
Unlike in previous studies (P+19 and L+20) where we used an average PSF in each band, we have accounted for the observed distribution of the PSFs in each band when creating the simulated arcs with simulated PSFs.

To demonstrate that this produces a clear advantage in the final completeness, 
we show in Fig.~\ref{fig:completeness_average_psf} the ROC (left panel) and completeness curves {(right panel)} that we obtain adopting {an average (thick lines) PSF and a variable one (thin lines).  The thin lines, colored in blue for the 3-band and red for the 1-band CNN, reproduce the ones shown in Figs.~\ref{fig:ROC_dist} and \ref{fig:completeness_real_data}. 
Interestingly, the PSF does not produce significant differences in terms of TPR/FPR (i.e. contamination). The same happens for the completeness on simulated lenses, for both CNNs.  
In other words, the CNNs ``get out what is put in''.
On real HQ lens candidates, the adoption of the average PSFs does not change significantly the completeness of the 1-band CNNs because 
the KiDS $r$-band images have a narrow {seeing} distribution. However, it heavily reduces the completeness of the 3-band CNN, because of the wider distribution of the real PSF of the images.}

This test gives a visual idea of the impact of using an appropriate PSF when producing $P_{\rm CNN}$ predictions. 
However, the step of accounting for a realistic PSF variation has been done only for the sources but not yet for the lenses,
which would impact the final completeness of real candidates.
Ultimately, to improve the realism of the simulated systems, one should use the same PSF for the arcs and deflectors while, when making predictions for each individual lens, one should take into account the observed PSF of each system.  
Neither of these aspects have been included yet in the CNNs presented here and, together with the color optimization discussed above, represent the next developments of our work. 
In particular, we plan to accurately model the PSFs in different bands from stars close to the galaxies belonging to the predictive samples. These will be input in a new CNN that has been trained with pairs of images including the simulated arcs and the corresponding PSFs.
Then, during the classification, the CNN will be able to weight the probability of an object to be a lens using appropriate information about the PSFs.

Finally, we list other possible sources of bias in the 
lens simulation, that might further reduce the predictive ability of our CNNs:
1) the distributions of the lens mass parameters (e.g. Einstein radius, axis ratios) and source parameters (magnitudes, colors, etc). If these 
do not
cover the {entire } real parameter space, the predictive power of the CNNs on outliers will be affected (e.g. lenses with Einstein radius smaller than 1''or larger than ''); 
2) the assumptions of the SIE profile and the ``single lens'' can be too simple. Some real lenses will have more than one foreground galaxies, so that both the light profile and the gravitational potential of the system will be different from the 
one obtained from a single galaxy. Although we have many candidates in cluster environments and pairs of galaxies in the current HQ sample, we cannot be sure that
the completeness of these particular classes of system is optimized by the current CNNs. Some lenses with multi-deflectors could be missed.

\section{Conclusions}
We have built two new CNN classifiers to search for strong gravitational lenses in both $1$-band {($r$)} and $3$-band ($gri$) color-combined images in {the } KiDS internal Data Release 5. The classifiers have been tested on both simulated data, made of real KiDS galaxies with artificial arcs following an empirical color distribution, and HQ candidates identified in previous KiDS releases. 

We have found different performances in terms of completeness of the recovered lenses when using mock data and real HQ candidates. {However, this difference is much smaller for real data, especially at higher threshold probabilities ($P_{\rm cnn}>0.7$). }
On simulated data, the $3$-band classifier performs much more efficiently (e.g. it reaches $\sim 95\%$ above a CNN probability $P_{\rm CNN}=0.9$) than the $1$-band classifier (which reaches $\sim 88\%$ at the same $P_{\rm CNN}=0.9$). On 
real data, the difference is almost cancelled because of a strong degradation of the $3$-band CNN performances. This has been mainly tracked to a combination of a {too blue }
color definition of the arcs, and the impact of the non-uniform seeing in different {KiDS } bands. 

We have demonstrated the presence of a clear complementarity between the 1-band and the 3-band CNNs, residing on the different 
features they are able to detect, at least in ground-based observations with sparse seeing distribution among the different filters. The 1-band focuses on the morphology of the strong lensing features, the 3-band on the color contrast between the red lens and the bluer arcs or multiple images. As a consequence, some lenses receiving a very high score from the 3-band CNNs might obtain a low probability for the 1-band one, and vice-versa.
For this reason, {at least for one of the ``predictive samples'', } we have decided to take advantage of the best capabilities of both classifiers to optimize the completeness and collect the best candidates selected from the two CNNs separately.

We have applied the new classifiers to two ``predictive samples'' selected from KiDS-iDR5: the luminous red galaxies (LRGs) and the bright galaxies (BGs). Using as probability thresholds $P_1>0.80$ ``or" $P_3>0.80$ for the 1-band and 3-band CNN respectively on the LRGs and a more conservative combination of $P_1>0.90$ ``and" $P_3>0.90$ for the BGs, the classifiers
have retrieved a total of 5810 candidates.  After a first {visual } cleaning {carried out by a single inspector}, we have defined a ``good sample'' of 487 potential lenses{. These objects were further inspected by 7 co-authors. Setting a threshold on the mean score obtained by 5 of them, after removing the highest and lowest scores, we  }
finally collected a HQ sample of 97 lenses.
These HQ lens candidates show a variety of different arcs or point-like configurations around central galaxies{. Due }
to the high $P_{\rm CNN}$ and human score, these are ideal candidates for follow-up spectroscopical observations.  A catalog of 268 HQ candidates, including the 97 in these work and the ones found by others works in KIDS (\citealt{2017MNRAS.472.1129P, 2019MNRAS.484.3879P}, L+20),  is made available at this link \href {https://kids.strw.leidenuniv.nl/DR4/hqlenses.php}{https://kids.strw.leidenuniv.nl/DR4/hqlenses.php}.

We {have finally discussed }
the avenues to improve the current classifiers, related 1) to the color definition of SGL sources used to produce realistic arc/multiple image colors and 2) to the way the PSF is accounted for both for arc simulation and for lens classification. In particular, we plan to implement the PSF modeling in different bands from stars close to the galaxies belonging to the predictive samples{. This information will be used as }
an additional “label” to be input into the next generation of classifiers. We expect this to produce a significant leap in the accuracy of our methods and to optimize the number of HQ candidates in future
large sky surveys (Rubin/LSST, Euclid, CSST) {while minimizing the number of }
false positives.

\section*{Acknowledgements}
RL acknowledge support from China Postdoctoral Science foundation 2020M672935, and 2021T140773, RL also acknowledge Guangdong Basic and Applied Basic Research Foundation 2019A1515110286. NRN acknowledge financial support from the “One hundred top talent program of Sun Yat-sen University” grant N. 71000-18841229.  NRN also acknowledge support from the European Union Horizon 2020 research and innovation programme under the Marie Skodowska-Curie grant agreement n. 721463 to the SUNDIAL ITN network. CS is supported by an `Hintze Fellow' at the Oxford Centre for Astrophysical Surveys, which is funded through generous support from the Hintze Family Charitable Foundation. GV has received funding from the European Union's Horizon 2020 research and innovation programme under the Marie Sklodovska-Curie grant agreement No 897124. AD acknowledge support from ERC Consolidator Grant (No. 770935). AHW is supported by an European Research Council Consolidator Grant (No. 770935). CH acknowledges support from the European Research Council under grant number 647112, and support from the Max Planck Society and the Alexander von Humboldt Foundation in the framework of the Max Planck-Humboldt Research Award endowed by the Federal Ministry of Education and Research. H. Hildebrandt is supported by a Heisenberg grant of the Deutsche Forschungsgemeinschaft (Hi 1495/5-1) as well as an ERC Consolidator Grant (No. 770935).

\begin{figure*}[htbp]
\centerline{\includegraphics[width=16cm]{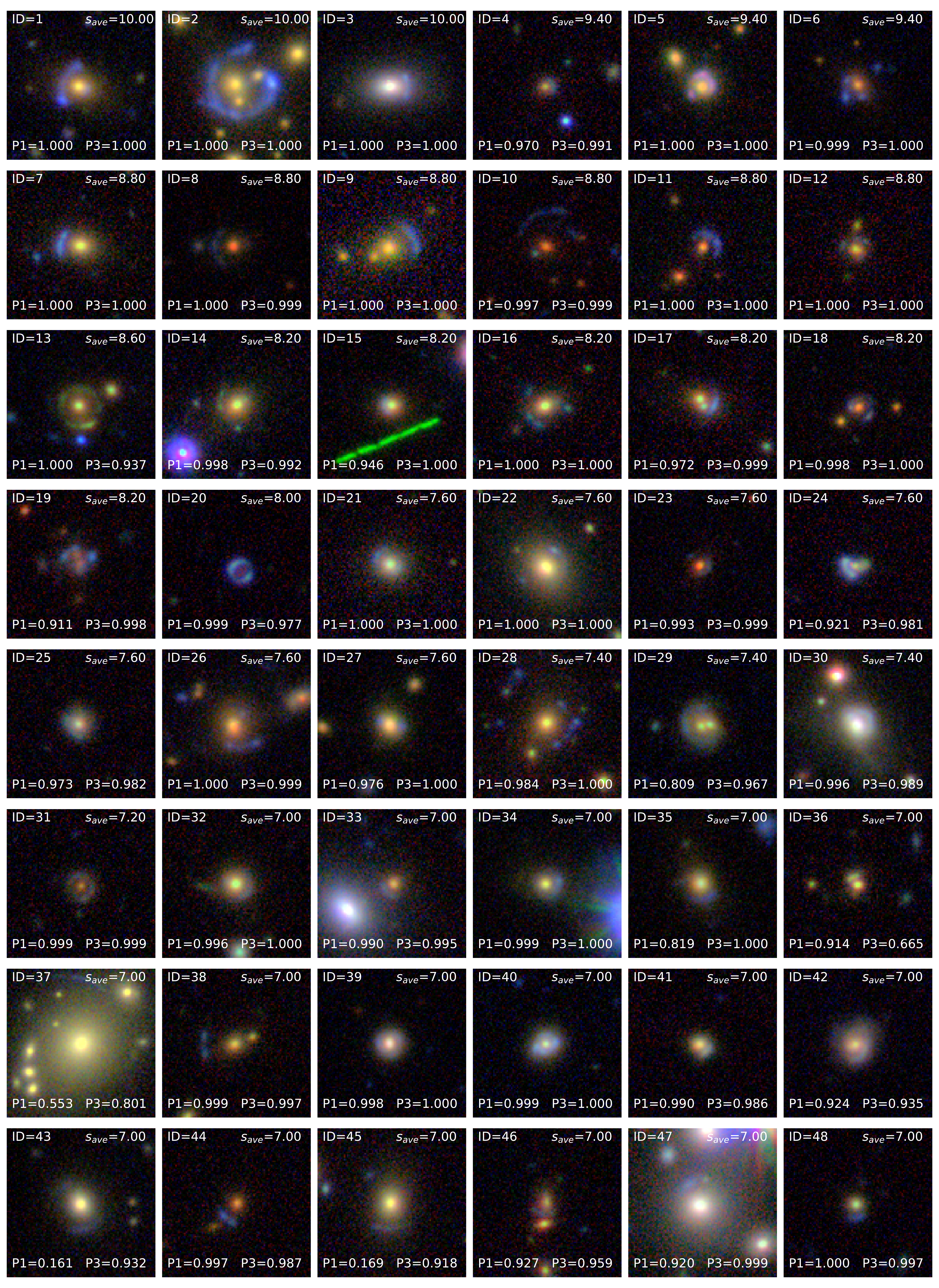}}
\caption{Colored stamps of the best candidates The stamps (20''x20'') are obtained by combining {\it g, r} and {\it i} KiDS images.}
\label{fig:high_quality_candidates}
\vspace{0.5cm}
\end{figure*}

\addtocounter{figure}{-1}
\begin{figure*}[htbp]
\centerline{\includegraphics[width=16cm]{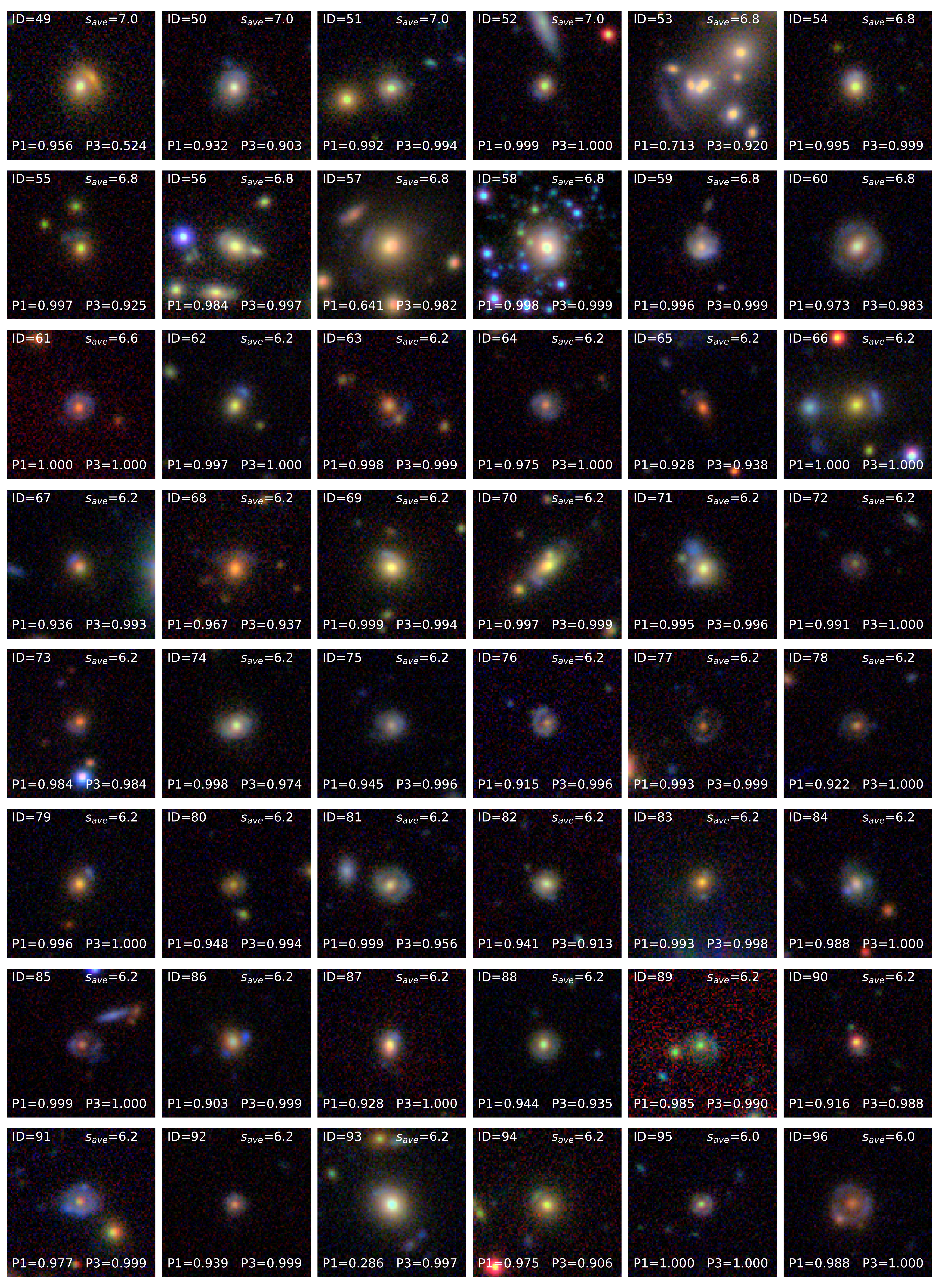}}
\caption{Continued}
\label{fig:high_quality_candidates1}
\vspace{0.5cm}
\end{figure*}

\addtocounter{figure}{-1}
\begin{figure*}[htbp]
\centerline{\includegraphics[width=2.9cm]{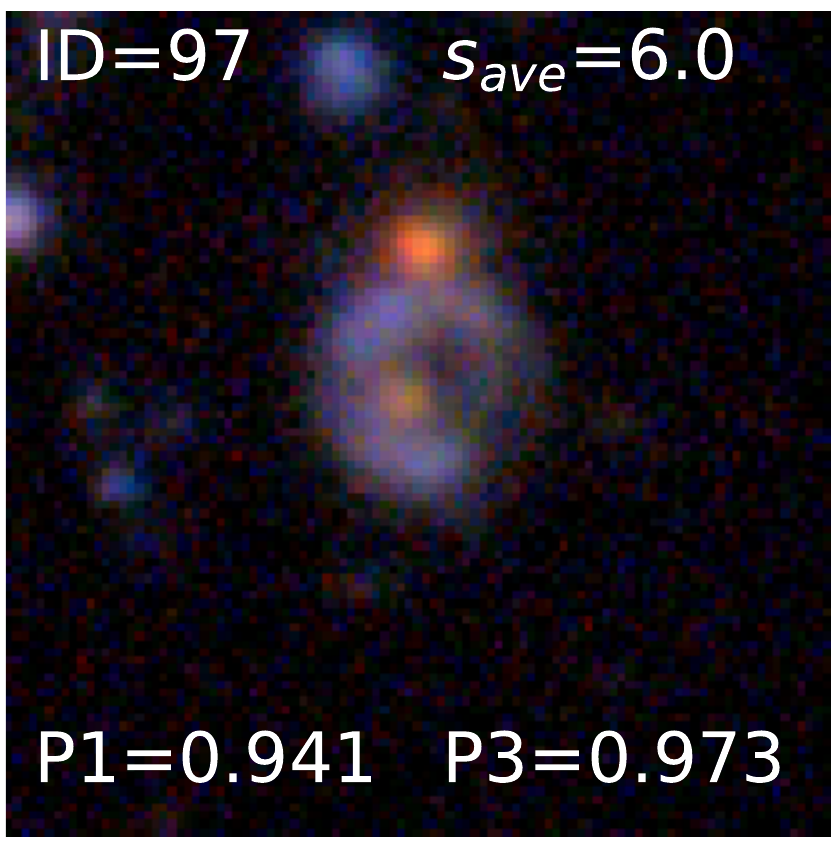}}
\caption{Continued}
\label{fig:high_quality_candidates1}
\vspace{0.5cm}
\end{figure*}

\begin{table*}[htbp]
\begin{center}
\caption{\label{tb:lens_candidates} Properties of the best 97 lens candidates} 
\begin{tabular}{l l l l l l l l l l l c }
\hline \hline
ID & KiDS$\_$ID & RAJ2000 & DECJ2000 & $ r_{\rm{auto}}$ & $P_1$ & $P_3$ & $s_{\rm ave}$ & RMS\\
\hline
1 & J125808.27$+$033208.86 & 194.53446 & 3.53579 & 18.4582 & 1.000 & 1.000 & 10.00 & 0.00 \\
2 & J020706.67$-$272644.71 & 31.77778 & $-$27.44575 & 17.6572 & 1.000 & 1.000 & 10.00 & 0.00 \\
3 & J124727.77$+$035701.16 & 191.86571 & 3.95032 & 17.6281 & 1.000 & 1.000 & 10.00 & 0.00 \\
4 & J124350.11$+$031434.75 & 190.95879 & 3.24299 & 20.4942 & 0.970 & 0.991 & 9.40 & 1.20 \\
5 & J111826.56$-$033739.06 & 169.61066 & $-$3.62752 & 18.5258 & 1.000 & 1.000 & 9.40 & 1.20 \\
6 & J234804.98$-$302855.45 & 357.02074 & $-$30.48207 & 19.8429 & 0.999 & 1.000 & 9.40 & 1.20 \\
7 & J124952.79$+$030933.21 & 192.46998 & 3.15923 & 18.7408 & 1.000 & 1.000 & 8.80 & 1.47 \\
8 & J020317.64$-$353056.15 & 30.82351 & $-$35.5156 & 20.4508 & 1.000 & 0.999 & 8.80 & 1.47 \\
9 & J020526.09$-$353947.36 & 31.35873 & $-$35.66315 & 18.8324 & 1.000 & 1.000 & 8.80 & 1.47 \\
10 & J133134.96$+$033139.35 & 202.89568 & 3.5276 & 20.805 & 0.997 & 0.999 & 8.80 & 1.47 \\
11 & J142822.48$+$031800.06 & 217.09365 & 3.30002 & 20.5881 & 1.000 & 1.000 & 8.80 & 1.47 \\
12 & J130004.86$+$014450.56 & 195.02023 & 1.74738 & 20.0374 & 1.000 & 1.000 & 8.80 & 1.47 \\
13 & J104724.45$-$031242.50 & 161.85186 & $-$3.21181 & 19.3084 & 1.000 & 0.937 & 8.60 & 2.80 \\
14 & J151737.38$+$024235.06 & 229.40577 & 2.70974 & 18.5053 & 0.998 & 0.992 & 8.20 & 1.47 \\
15 & J105559.06$+$022249.79 & 163.99607 & 2.3805 & 19.0038 & 0.946 & 1.000 & 8.20 & 1.47 \\
16 & J130526.14$-$035438.01 & 196.35892 & $-$3.91056 & 18.9711 & 1.000 & 1.000 & 8.20 & 1.47 \\
17 & J125505.51$-$032459.20 & 193.77294 & $-$3.41644 & 20.8538 & 0.972 & 0.999 & 8.20 & 1.47 \\
18 & J143354.72$+$030014.06 & 218.478 & 3.0039 & 20.474 & 0.998 & 1.000 & 8.20 & 1.47 \\
19 & J014907.97$-$313738.93 & 27.28322 & $-$31.62748 & 20.953 & 0.911 & 0.998 & 8.20 & 1.47 \\
20 & J023341.29$-$344759.24 & 38.42204 & $-$34.79979 & 20.7505 & 0.999 & 0.977 & 8.00 & 2.76 \\
21 & J120143.00$-$035815.58 & 180.42918 & $-$3.97099 & 18.589 & 1.000 & 1.000 & 7.60 & 1.20 \\
22 & J134404.67$-$030629.89 & 206.01944 & $-$3.1083 & 17.5325 & 1.000 & 1.000 & 7.60 & 1.20 \\
23 & J143018.63$+$034441.30 & 217.57761 & 3.74481 & 20.8596 & 0.993 & 0.999 & 7.60 & 1.20 \\
24 & J115351.61$-$033558.18 & 178.46503 & $-$3.59949 & 19.534 & 0.921 & 0.981 & 7.60 & 1.20 \\
25 & J121631.22$-$035304.15 & 184.13008 & $-$3.88449 & 19.3739 & 0.973 & 0.982 & 7.60 & 1.20 \\
26 & J224120.79$-$272937.60 & 340.33662 & $-$27.49378 & 18.5693 & 1.000 & 0.999 & 7.60 & 1.20 \\
27 & J014319.43$-$321914.54 & 25.83097 & $-$32.32071 & 18.9009 & 0.976 & 1.000 & 7.60 & 1.20 \\
28 & J124157.98$+$034721.94 & 190.49158 & 3.78943 & 17.9694 & 0.984 & 1.000 & 7.40 & 2.58 \\
29 & J112610.75$+$033847.89 & 171.54477 & 3.64664 & 19.4704 & 0.809 & 0.967 & 7.40 & 2.58 \\
30 & J122203.54$+$030743.20 & 185.51474 & 3.12867 & 17.351 & 0.996 & 0.989 & 7.40 & 2.58 \\
31 & J115247.36$+$034255.18 & 178.19734 & 3.71533 & 20.3757 & 0.999 & 0.999 & 7.20 & 3.43 \\
32 & J003639.79$-$342535.61 & 9.16581 & $-$34.42656 & 18.7459 & 0.996 & 1.000 & 7.00 & 0.00 \\
33 & J130021.54$-$035610.71 & 195.08977 & $-$3.93631 & 19.2234 & 0.990 & 0.995 & 7.00 & 0.00 \\
34 & J001011.07$-$270046.47 & 2.54611 & $-$27.01291 & 17.433 & 0.999 & 1.000 & 7.00 & 0.00 \\
35 & J014658.63$-$264616.68 & 26.74428 & $-$26.7713 & 18.978 & 0.819 & 1.000 & 7.00 & 0.00 \\
36 & J124100.73$+$031802.65 & 190.25303 & 3.30074 & 19.6761 & 0.914 & 0.665 & 7.00 & 0.00 \\
37 & J143821.90$+$034013.28 & 219.59123 & 3.67036 & 16.3782 & 0.553 & 0.801 & 7.00 & 0.00 \\
38 & J004919.61$-$352730.58 & 12.33172 & $-$35.4585 & 19.8968 & 0.999 & 0.997 & 7.00 & 0.00 \\
39 & J231852.22$-$273759.16 & 349.71757 & $-$27.6331 & 18.9433 & 0.998 & 1.000 & 7.00 & 0.00 \\
40 & J145605.71$+$033707.02 & 224.0238 & 3.61862 & 18.9394 & 0.999 & 1.000 & 7.00 & 0.00 \\
41 & J004419.36$-$350407.80 & 11.08066 & $-$35.06883 & 19.5136 & 0.990 & 0.986 & 7.00 & 0.00 \\
42 & J230502.92$-$264517.15 & 346.26215 & $-$26.75476 & 18.7199 & 0.924 & 0.935 & 7.00 & 0.00 \\
43 & J120813.61$+$035444.17 & 182.0567 & 3.91227 & 18.301 & 0.161 & 0.932 & 7.00 & 0.00 \\
44 & J004838.52$-$351355.91 & 12.1605 & $-$35.2322 & 20.5841 & 0.997 & 0.987 & 7.00 & 0.00 \\
45 & J122455.77$+$031337.92 & 186.23238 & 3.2272 & 18.2486 & 0.169 & 0.918 & 7.00 & 0.00 \\
46 & J123129.73$+$034716.39 & 187.87387 & 3.78789 & 20.1557 & 0.927 & 0.959 & 7.00 & 0.00 \\
47 & J133221.82$-$031517.42 & 203.0909 & $-$3.25484 & 16.5788 & 0.920 & 0.999 & 7.00 & 0.00 \\
48 & J014538.40$-$305936.15 & 26.40999 & $-$30.99338 & 20.1657 & 1.000 & 0.997 & 7.00 & 0.00 \\
49 & J103244.87$+$035608.11 & 158.18696 & 3.93559 & 18.5431 & 0.956 & 0.524 & 7.00 & 0.00 \\
50 & J132731.02$+$033655.73 & 201.87927 & 3.61548 & 18.8195 & 0.932 & 0.903 & 7.00 & 0.00 \\
51 & J122917.55$+$031330.57 & 187.32312 & 3.22516 & 19.0799 & 0.992 & 0.994 & 7.00 & 0.00 \\
52 & J120355.23$-$033218.89 & 180.98013 & $-$3.53858 & 19.8452 & 0.999 & 1.000 & 7.00 & 0.00 \\
53 & J015148.40$-$323715.86 & 27.95167 & $-$32.62107 & 17.9265 & 0.713 & 0.920 & 6.80 & 2.23 \\
54 & J125632.39$-$021600.64 & 194.13498 & $-$2.26684 & 19.0053 & 0.995 & 0.999 & 6.80 & 2.23 \\
55 & J013336.28$-$321114.42 & 23.40117 & $-$32.18734 & 19.8008 & 0.997 & 0.925 & 6.80 & 2.23 \\
56 & J134223.83$-$035645.69 & 205.59929 & $-$3.94603 & 18.5063 & 0.984 & 0.997 & 6.80 & 2.23 \\
57 & J014320.63$-$271746.99 & 25.83596 & $-$27.29639 & 17.6511 & 0.641 & 0.982 & 6.80 & 2.23 \\
58 & J024000.80$-$341812.44 & 40.00335 & $-$34.30346 & 17.6939 & 0.998 & 0.999 & 6.80 & 2.23 \\
\hline
\hline
\end{tabular}
\end{center}
\textsc{      Note.} --- We list from Column 1 to 4, the ID, the KiDS name and the coordinates (in degrees) of the candidates, respectively. 
Column 5 lists the total magnitudes ($r_{\rm auto}$) obtained by from SExtractor. 
Column 6 and 7 list the probabily to be a lens from 1-band CNN and 3-band CNN. 
Column 8 and 9 list the average scores from human inspection and the corresponding RMS.).
\end{table*}
\addtocounter{table}{-1}

\begin{table*}[htbp]
\begin{center}
\caption{\label{tb:tb1} Properties of the best 82 lens candidates} 
\begin{tabular}{l l l l l l l l l l l c }
\hline \hline
ID & KiDS$\_$ID & RAJ2000 & DECJ2000 & $ r_{\rm{auto}}$ & $P_1$ & $P_3$ & $s_{\rm ave}$ & RMS\\
\hline
59 & J111401.10$-$033323.54 & 168.50457 & $-$3.55654 & 19.0885 & 0.996 & 0.999 & 6.80 & 2.23 \\
60 & J222201.35$-$273614.23 & 335.50565 & $-$27.60395 & 18.7133 & 0.973 & 0.983 & 6.80 & 2.23 \\
61 & J023206.32$-$352923.41 & 38.02635 & $-$35.48984 & 20.0357 & 1.000 & 1.000 & 6.60 & 3.14 \\
62 & J000651.87$-$271357.05 & 1.71611 & $-$27.23251 & 19.591 & 0.997 & 1.000 & 6.20 & 1.60 \\
63 & J031418.60$-$284156.45 & 48.57751 & $-$28.69901 & 20.0205 & 0.998 & 0.999 & 6.20 & 1.60 \\
64 & J003118.07$-$350132.27 & 7.82531 & $-$35.02563 & 20.0741 & 0.975 & 1.000 & 6.20 & 1.60 \\
65 & J102839.00$-$030048.13 & 157.16251 & $-$3.01337 & 20.6399 & 0.928 & 0.938 & 6.20 & 1.60 \\
66 & J234124.17$-$271145.14 & 355.35073 & $-$27.19587 & 19.1013 & 1.000 & 1.000 & 6.20 & 1.60 \\
67 & J023517.89$-$271129.01 & 38.82454 & $-$27.19139 & 17.9813 & 0.936 & 0.993 & 6.20 & 1.60 \\
68 & J002342.53$-$350534.93 & 5.9272 & $-$35.09304 & 19.3053 & 0.967 & 0.937 & 6.20 & 1.60 \\
69 & J125940.84$-$032637.29 & 194.92017 & $-$3.44369 & 18.418 & 0.999 & 0.994 & 6.20 & 1.60 \\
70 & J020518.38$-$273743.38 & 31.32657 & $-$27.62872 & 18.6231 & 0.997 & 0.999 & 6.20 & 1.60 \\
71 & J014535.28$-$313235.48 & 26.39699 & $-$31.54319 & 18.7319 & 0.995 & 0.996 & 6.20 & 1.60 \\
72 & J003617.62$-$343420.56 & 9.07342 & $-$34.57238 & 20.9532 & 0.991 & 1.000 & 6.20 & 1.60 \\
73 & J021203.87$-$270653.29 & 33.01613 & $-$27.1148 & 20.2011 & 0.984 & 0.984 & 6.20 & 1.60 \\
74 & J000654.69$-$283746.64 & 1.72787 & $-$28.62962 & 19.1697 & 0.998 & 0.974 & 6.20 & 1.60 \\
75 & J002823.72$-$265957.95 & 7.09883 & $-$26.99943 & 19.6856 & 0.945 & 0.996 & 6.20 & 1.60 \\
76 & J134604.91$-$032322.36 & 206.52046 & $-$3.38954 & 20.5645 & 0.915 & 0.996 & 6.20 & 1.60 \\
77 & J132205.60$-$014938.03 & 200.52334 & $-$1.82723 & 20.7019 & 0.993 & 0.999 & 6.20 & 1.60 \\
78 & J132326.60$+$033701.12 & 200.86085 & 3.61698 & 20.5082 & 0.922 & 1.000 & 6.20 & 1.60 \\
79 & J120913.95$+$031014.33 & 182.30813 & 3.17065 & 19.9832 & 0.996 & 1.000 & 6.20 & 1.60 \\
80 & J110322.38$-$010726.73 & 165.84325 & $-$1.12409 & 20.5388 & 0.948 & 0.994 & 6.20 & 1.60 \\
81 & J112307.43$-$032628.80 & 170.78097 & $-$3.44133 & 19.3007 & 0.999 & 0.956 & 6.20 & 1.60 \\
82 & J110256.69$-$011831.10 & 165.73622 & $-$1.30864 & 19.4852 & 0.941 & 0.913 & 6.20 & 1.60 \\
83 & J124914.12$-$033404.02 & 192.30883 & $-$3.56778 & 18.814 & 0.993 & 0.998 & 6.20 & 1.60 \\
84 & J014403.02$-$321237.07 & 26.01258 & $-$32.2103 & 19.7753 & 0.988 & 1.000 & 6.20 & 1.60 \\
85 & J012246.67$-$273322.09 & 20.69444 & $-$27.55613 & 20.8031 & 0.999 & 1.000 & 6.20 & 1.60 \\
86 & J014603.78$-$264457.39 & 26.51576 & $-$26.74927 & 19.7145 & 0.903 & 0.999 & 6.20 & 1.60 \\
87 & J111500.11$-$025804.30 & 168.75047 & $-$2.96786 & 19.2168 & 0.928 & 1.000 & 6.20 & 1.60 \\
88 & J143738.88$-$031635.12 & 219.41199 & $-$3.27642 & 19.7014 & 0.944 & 0.935 & 6.20 & 1.60 \\
89 & J133214.79$-$005719.17 & 203.06162 & $-$0.95533 & 19.8245 & 0.985 & 0.990 & 6.20 & 1.60 \\
90 & J125935.77$-$024708.16 & 194.89906 & $-$2.7856 & 20.4277 & 0.916 & 0.988 & 6.20 & 1.60 \\
91 & J115735.71$+$034500.96 & 179.39877 & 3.75027 & 19.0632 & 0.977 & 0.999 & 6.20 & 1.60 \\
92 & J145209.65$+$034947.22 & 223.04023 & 3.82978 & 20.6841 & 0.939 & 0.999 & 6.20 & 1.60 \\
93 & J112440.94$+$030347.95 & 171.17057 & 3.06332 & 17.9943 & 0.286 & 0.997 & 6.20 & 1.60 \\
94 & J125148.11$+$034128.43 & 192.95046 & 3.69123 & 18.9286 & 0.975 & 0.906 & 6.20 & 1.60 \\
95 & J122932.16$+$033726.90 & 187.38402 & 3.62414 & 20.4791 & 1.000 & 1.000 & 6.00 & 2.68 \\
96 & J232142.07$-$271318.88 & 350.42531 & $-$27.22191 & 19.6246 & 0.988 & 1.000 & 6.00 & 2.68 \\
97 & J010552.76$-$352030.92 & 16.46984 & $-$35.34192 & 19.5707 & 0.941 & 0.973 & 6.00 & 3.95 \\
\hline
\hline
\end{tabular}
\end{center}
\textsc{      Note.} --- continued.
\end{table*}

\end{document}